\documentclass[ALICE,manyauthors]{cernphprep}
\usepackage[comma,square,numbers,sort&compress]{natbib}
\usepackage{hyperref}
\usepackage{lineno}
\usepackage{xspace}
\usepackage{color}
\usepackage{verbatim}
\usepackage{multirow}
\usepackage[T1]{fontenc}
\usepackage{orcidlink}

\begin{document}
%

\newcommand{\pp}           {pp\xspace}
\newcommand{\ppbar}        {\mbox{$\mathrm {p\overline{p}}$}\xspace}
\newcommand{\XeXe}         {\mbox{Xe--Xe}\xspace}
\newcommand{\PbPb}         {\mbox{Pb--Pb}\xspace}
\newcommand{\pA}           {\mbox{pA}\xspace}
\newcommand{\pPb}          {\mbox{p--Pb}\xspace}
\newcommand{\AuAu}         {\mbox{Au--Au}\xspace}
\newcommand{\dAu}          {\mbox{d--Au}\xspace}
\newcommand{\UU}         {\mbox{U--U}\xspace}
\newcommand{\OO}         {\mbox{O--O}\xspace}
\newcommand{\RuRu}         {\mbox{Ru--Ru}\xspace}
\newcommand{\ZrZr}         {\mbox{Zr--Zr}\xspace}

\newcommand{\s}            {\ensuremath{\sqrt{s}}\xspace}
\newcommand{\snn}          {\ensuremath{\sqrt{s_{\mathrm{NN}}}}\xspace}
\newcommand{\pt}           {\ensuremath{p_{\rm T}}\xspace}
\newcommand{\meanpt}       {$\langle p_{\mathrm{T}}\rangle$\xspace}
\newcommand{\ycms}         {\ensuremath{y_{\rm CMS}}\xspace}
\newcommand{\ylab}         {\ensuremath{y_{\rm lab}}\xspace}
\newcommand{\etarange}[1]  {\mbox{$\left | \eta \right |~<~#1$}}
\newcommand{\etagap}[1]  {\mbox{$\left | \Delta\eta \right | > #1$}}
\newcommand{\Etarange}[2]  {\mbox{$#1~<~ \eta ~<~#2$}}
\newcommand{\ptrange}[2]  {\mbox{$#1~<~ p_{\mathrm{T}} ~<~#2$\xspace Ge\kern-.1emV/$c$\xspace}}
\newcommand{\yrange}[1]    {\mbox{$\left | y \right |~<~#1$}}
\newcommand{\dndy}         {\ensuremath{\mathrm{d}N_\mathrm{ch}/\mathrm{d}y}\xspace}
\newcommand{\dndeta}       {\ensuremath{\mathrm{d}N_\mathrm{ch}/\mathrm{d}\eta}\xspace}
\newcommand{\avdndeta}     {\ensuremath{\langle\dndeta\rangle}\xspace}
\newcommand{\dNdy}         {\ensuremath{\mathrm{d}N_\mathrm{ch}/\mathrm{d}y}\xspace}
\newcommand{\Npart}        {\ensuremath{N_\mathrm{part}}\xspace}
\newcommand{\Ncoll}        {\ensuremath{N_\mathrm{coll}}\xspace}
\newcommand{\dEdx}         {\ensuremath{\textrm{d}E/\textrm{d}x}\xspace}
\newcommand{\RpPb}         {\ensuremath{R_{\rm pPb}}\xspace}

\newcommand{\nineH}        {$\sqrt{s}~=~0.9$~Te\kern-.1emV\xspace}
\newcommand{\seven}        {$\sqrt{s}~=~7$~Te\kern-.1emV\xspace}
\newcommand{\twoH}         {$\sqrt{s}~=~0.2$~Te\kern-.1emV\xspace}
\newcommand{\twosevensix}  {$\sqrt{s}~=~2.76$~Te\kern-.1emV\xspace}
\newcommand{\five}         {$\sqrt{s}~=~5.02$~Te\kern-.1emV\xspace}
\newcommand{\fivefourfour}         {$\sqrt{s_{_{\mathrm{NN}}}}~=~5.44$~Te\kern-.1emV\xspace}
\newcommand{\twosevensixnn}{$\sqrt{s_{\mathrm{NN}}}~=~2.76$~Te\kern-.1emV\xspace}
\newcommand{\fivenn}       {$\sqrt{s_{_{\mathrm{NN}}}}~=~5.02$~Te\kern-.1emV\xspace}
\newcommand{\LT}           {L{\'e}vy-Tsallis\xspace}
\newcommand{\GeVc}         {Ge\kern-.1emV/$c$\xspace}
\newcommand{\MeVc}         {Me\kern-.1emV/$c$\xspace}
\newcommand{\TeV}          {Te\kern-.1emV\xspace}
\newcommand{\GeV}          {Ge\kern-.1emV\xspace}
\newcommand{\MeV}          {Me\kern-.1emV\xspace}
\newcommand{\GeVmass}      {Ge\kern-.2emV/$c^2$\xspace}
\newcommand{\MeVmass}      {Me\kern-.2emV/$c^2$\xspace}
\newcommand{\lumi}         {\ensuremath{\mathcal{L}}\xspace}

\newcommand{\ITS}          {\rm{ITS}\xspace}
\newcommand{\TOF}          {\rm{TOF}\xspace}
\newcommand{\ZDC}          {\rm{ZDC}\xspace}
\newcommand{\ZDCs}         {\rm{ZDCs}\xspace}
\newcommand{\ZNA}          {\rm{ZNA}\xspace}
\newcommand{\ZNC}          {\rm{ZNC}\xspace}
\newcommand{\SPD}          {\rm{SPD}\xspace}
\newcommand{\SDD}          {\rm{SDD}\xspace}
\newcommand{\SSD}          {\rm{SSD}\xspace}
\newcommand{\TPC}          {\rm{TPC}\xspace}
\newcommand{\TRD}          {\rm{TRD}\xspace}
\newcommand{\VZERO}        {\rm{V0}\xspace}
\newcommand{\VZEROA}       {\rm{V0A}\xspace}
\newcommand{\VZEROC}       {\rm{V0C}\xspace}
\newcommand{\Vdecay} 	   {\ensuremath{V^{0}}\xspace}

\newcommand{\ee}           {\ensuremath{e^{+}e^{-}}} 
\newcommand{\pip}          {\ensuremath{\pi^{+}}\xspace}
\newcommand{\pim}          {\ensuremath{\pi^{-}}\xspace}
\newcommand{\kap}          {\ensuremath{\rm{K}^{+}}\xspace}
\newcommand{\kam}          {\ensuremath{\rm{K}^{-}}\xspace}
\newcommand{\pbar}         {\ensuremath{\rm\overline{p}}\xspace}
\newcommand{\kzero}        {\ensuremath{{\rm K}^{0}_{\rm{S}}}\xspace}
\newcommand{\lmb}          {\ensuremath{\Lambda}\xspace}
\newcommand{\almb}         {\ensuremath{\overline{\Lambda}}\xspace}
\newcommand{\Om}           {\ensuremath{\Omega^-}\xspace}
\newcommand{\Mo}           {\ensuremath{\overline{\Omega}^+}\xspace}
\newcommand{\X}            {\ensuremath{\Xi^-}\xspace}
\newcommand{\Ix}           {\ensuremath{\overline{\Xi}^+}\xspace}
\newcommand{\Xis}          {\ensuremath{\Xi^{\pm}}\xspace}
\newcommand{\Oms}          {\ensuremath{\Omega^{\pm}}\xspace}
\newcommand{\degree}       {\ensuremath{^{\rm o}}\xspace}

\begin{titlepage}
\PHyear{2024}       
\PHnumber{228}      
\PHdate{06 September}  

\title{Exploring nuclear structure with multiparticle azimuthal correlations at the LHC}
\ShortTitle{Exploring nuclear structure with flow measurements}   

\Collaboration{ALICE Collaboration\thanks{See Appendix~\ref{app:collab} for the list of collaboration members}}
\ShortAuthor{ALICE Collaboration} 

\begin{abstract}

Details of the nuclear structure of $^{\rm 129}$Xe, such as the quadrupole deformation and the nuclear diffuseness, are studied by extensive measurements of anisotropic-flow-related observables in Xe--Xe collisions at a centre-of-mass energy per nucleon pair $\sqrt{s_{_{\mathrm{NN}}}}~=~5.44$~TeV with the ALICE detector at the LHC. The results are compared with those from Pb--Pb collisions at $\sqrt{s_{_{\mathrm{NN}}}}~=~5.02$~TeV for a baseline, given that the $^{\rm 208}$Pb nucleus exhibits a very weak deformation. Furthermore, comprehensive comparisons are performed with a state-of-the-art hybrid model using IP-Glasma+MUSIC+UrQMD. It is found that among various IP-Glasma+MUSIC+UrQMD calculations with different values of nuclear parameters, the one using a nuclear diffuseness parameter of $a_0=0.492$ and a nuclear quadrupole deformation parameter of $\beta_2=0.207$ provides a better description of the presented flow measurements.
These studies represent the first systematic exploration of nuclear structure at TeV energies, utilizing a comprehensive set of anisotropic flow observables. The measurements serve as a critical experimental benchmark for rigorously testing the interplay between nuclear structure inputs and heavy-ion theoretical models.

\end{abstract}
\end{titlepage}

\setcounter{page}{2} 


\section{Introduction}
\label{sec:Introduction}
Over the past two decades, low-energy nuclear physics has made remarkable progress. Advancements in experimental methods such as laser spectroscopy and Coulomb excitation techniques reveal additional insights into the size and shape of atomic nuclei~\cite{Lu:2013ena, Heylen:2020cco, Bree:2014mxa, Ayangeakaa:2019psv, Koszorus:2020mgn, Warbinek:2024ncq}. On the theoretical side, the advent of \textit{ab-initio} methods has allowed the description of light and medium-mass nuclei from first principles~\cite{Hergert:2020bxy,Gandolfi:2020pbj,Soma:2020xhv,Lahde:2019npb,Ekstrom:2022yea} and a flagship calculation of $^{208}$Pb has been recently reported~\cite{Hu:2021trw}. Nevertheless, systematic calculations of heavy-mass systems are still not yet possible, in particular, due to the computational difficulty in handling the (necessary) three-body nuclear interaction in large model spaces~\cite{Miyagi:2021pdc}.
Recent studies in high-energy heavy-ion collisions at the Relativistic Heavy-Ion Collider (RHIC)~\cite{STAR:2015mki,STAR:2021mii, Zhang:2021kxj,STAR:2024wgy,Zhao:2024feh} and the Large Hadron Collider (LHC)~\cite{ALICE:2018lao,ALICE:2018yvr,ALICE:2018cpu,ATLAS:2022dov,CMS:2019cyz} have demonstrated that nuclear collisions at ultrarelativistic energies offer promising new approaches for nuclear structure studies.
These studies successfully probed the nuclear shape from light to heavy nuclei~\cite{Giacalone:2021udy, Zhang:2021kxj,STAR:2024wgy,Zhao:2024feh,Zhao:2022uhl,ALICE:2021gxt,Bally:2021qys,Xu:2024bdh,Ryssens:2023fkv} and the neutron skin of $^{208}$Pb, $^{96}$Zr, and $^{96}$Ru~\cite{Giacalone:2023cet,Li:2019kkh}. 
Among these experimental approaches, anisotropic flow phenomena have been found to carry the imaging power of the nuclear structures at relativistic energies~\cite{Zhang:2021kxj,Xu:2021uar,Jia:2021qyu,Jia:2021tzt,Giacalone:2021udy,Magdy:2022cvt,Nielsen:2023znu,Lu:2023fqd,Zhao:2024lpc}. 
Anisotropic flow, which quantifies the anisotropic azimuthal distribution of the momenta of the produced particles, reflects the initial geometry and fluctuations of the overlapping region and probes the shape (or structure) of the colliding nuclei~\cite{Muller:2012zq,Drescher:2007cd,Heinz:2013th,Molnar:2001ux,Song:2017wtw,ALICE:2022wpn}.
The anisotropic flow is characterised by the Fourier expansion of the azimuthal distribution of produced particles~\cite{Voloshin:1994mz}
\begin{equation}
    \frac{dN}{d\varphi} \propto 1+2\sum\limits_{n=1}\limits^{\infty}{v_n\cos[n(\varphi-\Psi_n)]},
    \label{eq:FourierSeries}
\end{equation}
where $\varphi$ is the azimuthal angle of particle momentum and $\Psi_n$ is the $n^{\rm th}$-order symmetry plane.
The coefficients $v_n$ are called flow coefficients and can be calculated as
\begin{equation}
    v_n = \left\langle\cos[n(\varphi-\Psi_n)]\right\rangle.
    \label{eq:FlowCoefficient}
\end{equation}
Here, the brackets $\left\langle\right\rangle$ denote an average over all particles in one event.
With $v_n$ and $\Psi_n$, the n$^{\rm th}$ order (complex) anisotropic flow $V_n$ are defined as
\begin{equation}
V_n \equiv v_n \mathrm{e}^{i n\Psi_n}.
\end{equation} 

Systematic measurements of $v_n$~\cite{ALICE:2011ab,ALICE:2016ccg, ALICE:2018lao, ALICE:2019zfl, ATLAS:2012at,ATLAS:2019dct, CMS:2013wjq,CMS:2019cyz,STAR:2015mki}, event-by-event flow fluctuations~\cite{ALICE:2018rtz, ALICE:2022dtx,ALICE:2024fcv,ALICE:2023tvh,ATLAS:2013xzf,CMS:2017glf}, and correlations between various flow coefficients~\cite{ALICE:2016kpq, ALICE:2017fcd, ALICE:2017kwu, ALICE:2020sup, ALICE:2021adw, ATLAS:2015qwl} enabled the extraction of the transport properties of the Quark-Gluon Plasma (QGP) and to constrain the initial conditions of the heavy-ion collisions~\cite{Parkkila:2021tqq}. It has been shown that the low-harmonic flow coefficients are linearly correlated with the initial eccentricity coefficients of the same order~\cite{Niemi:2012aj,Song:2010mg} and that the higher harmonic flow coefficients, in particular their nonlinear flow mode, carry information about the correlations between different participant planes~\cite{ALICE:2017fcd, ALICE:2020sup}.
Furthermore, the correlation between $v_{2}$ and $v_{3}$, characterised by normalised symmetric cumulants NSC$(3,2)$~\cite{Bilandzic:2013kga}, has been found to reflect correlation between $\varepsilon_2$ and $\varepsilon_3$ eccentricity coefficients~\cite{Zhu:2016puf,ALICE:2016kpq,ALICE:2021adw}. These observables are widely recognised as powerful tools for precisely constraining the initial conditions of relativistic heavy-ion collisions~\cite{ALICE:2022wpn}.

For the initial state of heavy-ion collisions, the nuclear density profile $\rho(r,\theta,\phi)$ of the colliding nuclei can be described by the Woods--Saxon distribution~\cite{Hagino:2006fj,Jia:2021tzt}
\begin{equation}
\begin{split}
    \rho(r,\theta,\phi) = \frac{\rho_0}{1+e^{[r-R(\theta,\phi)]/a_0}},
    \label{eq:WoodsSaxon}
\end{split}
\end{equation}
where $r$, $\theta$, and $\phi$ define the position of a nucleon presented in spherical coordinates, of which the origin is the centre of the nucleus. The constant $\rho_0$ ensures that the integral of the distribution corresponds to the number of nucleons in the nucleus. The $a_0$ parameter represents the nuclear diffuseness. The $R(\theta,\phi) = R_0(1+\beta_2[\cos\gamma Y_{2,0}+\sin\gamma Y_{2,2}])$ term models the nuclear surface expanded in terms of spherical harmonics $Y_{n,m}$, keeping terms up to $n=2$ that are the most relevant in the structure of $^{129}$Xe~\cite{Jia:2021qyu,Bally:2021qys,ALICE:2018yvr}.
Notably, $Y_{2,-2}$, $Y_{2,-1}$, and $Y_{2,1}$ are utilised to establish the intrinsic frame, which renders $Y_{2,0}$ and $Y_{2,2}$ as the only pertinent degrees of freedom.
In $R(\theta,\phi)$, $R_0$ denotes the nuclear radius, and $\beta_2$ is the quadrupole deformation parameter.
In low-energy nuclear experiments, $\beta_2$ for even-A isotopes of Xe can be extracted using the electric quadrupole transition probability B(E2)$\uparrow$ from the ground 0$^{+}$ to the first-excited 2$^{+}$ state~\cite{Raman:2001nnq,Pritychenko:2013gwa}, although such extraction can be deficient by approximately 20\% due to fragmentation of the low-lying electric-quadrupole strength~\cite{Henderson:2020yql}.
By interpolating the values between $^{\rm 128}$Xe and $^{\rm 130}$Xe, $\beta_2$ for $^{\rm 129}$Xe was estimated to be $0.18 \pm 0.02$~\cite{ALICE:2018yvr}. 
Finally, the triaxial parameter $\gamma$ reflects the inequality of the axes of the spheroid.

As described flow observables effectively capture a snapshot of the initial geometry of the collision and, by extension, offer a glimpse into the structure of the colliding nuclei, such as quadrupole deformation and triaxial structure. This ``imaging power'' of complex flow observables has been validated in recent theoretical model calculations and has shown great promise~\cite{Jia:2021qyu,Bally:2021qys,Zhao:2022uhl,Lu:2023fqd,Ryssens:2023fkv,Zhao:2024lpc}. A systematic study of various anisotropic flow observables is essential for investigating nuclear structure at ultrarelativistic energies. Nevertheless, only simple flow observables involving fewer particle correlations, such as $v_{n}$ coefficients, have been measured and used for studying nuclear structure~\cite{ALICE:2018lao,STAR:2024wgy}. The remaining, more complex flow observables, which involve multiparticle correlations and are likely more sensitive to the structure of the colliding nuclei~\cite{Jia:2022ozr,Lu:2023fqd}, have not yet been explored experimentally.

This Letter presents systematic measurements of a comprehensive set of flow observables using charged particles from \XeXe collisions at a centre-of-mass energy per nucleon pair \fivefourfour recorded by the ALICE detector, representing their first application to probe nuclear structure in heavy-ion collisions.
In addition, the corresponding measurements from \PbPb collisions at \fivenn, which provide a baseline because of the near-spherical shape of $^{\rm 208}$Pb~\cite{Raman:2001nnq}, are shown. 
Observables used in this study, including flow coefficients, flow fluctuations, nonlinear flow modes, and correlations between flow coefficients, are introduced in Sec.~\ref{sec:Observables}.
Section~\ref{sec:AnalysisDetail} presents the experimental setup and the evaluation of systematical uncertainties. The results are discussed in Sec.~\ref{sec:Results}, followed by the summary in Sec.~\ref{sec:Summary}.

\section{Observables and analysis method}
\label{sec:Observables}

Flow coefficients $v_n$ are usually measured by using two and four-particle cumulants~\cite{Bilandzic:2013kga,Bilandzic:2010jr,Borghini:2000sa,Moravcova:2020wnf}
\begin{equation} 
    \begin{split}
    v_n\{2\} &\equiv \sqrt{c_n\{2\}}, \\
    v_n\{4\} &\equiv\sqrt[4]{-c_n\{4\}},
    \label{eq:TwoParticleCorrelation}
     \end{split}
\end{equation}
where $c_n\{2\}$ and $c_n\{4\}$ are the two and four-particle cumulants, respectively.
It is known that $v_n\{2\}$ and $v_n\{4\}$ carry opposite contributions from flow fluctuations to the cumulant estimates~\cite{Voloshin:2007pc}.
When non-flow effects, which are the azimuthal angle correlations not associated with the symmetry plane, are small, the flow coefficients can be split into mean flow and flow fluctuation according to
\begin{equation}
    \begin{split}
    v_n\{2\}^2&\approx\langle v_n \rangle^2+\sigma_{v_n}^2,\\
    v_n\{4\}^2&\approx\langle v_n \rangle^2-\sigma_{v_n}^2.
    \label{eq:FlowFluctuations}
    \end{split}
\end{equation}
Here $\sigma_{v_n}$ is the standard deviation of the $v_n$ distribution, known as event-by-event fluctuation of $v_n$, and $\langle v_n \rangle$ is the mean value of the $v_n$ distribution.

For $n=2$ and $n=3$, $v_n$ coefficients for central and midcentral collisions are linearly correlated with the initial anisotropy coefficients $\varepsilon_n$~\cite{Niemi:2012aj,Song:2010mg}, where $\varepsilon_n$ is determined from the initial energy density profile~\cite{Alver:2010gr}
\begin{equation}
    \varepsilon_n e^{in\Phi_n} = -\frac{\langle r^n e^{in\phi} \rangle}{\langle r^n \rangle}~(n>1),
    \label{Eccentricity}
\end{equation}
where $\langle \rangle$ represents an average among the transverse positions $(r,\phi)$ of all participating nucleons, with $\phi$ representing the azimuthal angle and $r$ characterising the radial distance from the origin of the system. The $\Phi_n$ angle defines the symmetry plane of participant nucleons in the initial conditions.
Recent studies have shown that nuclear quadrupole deformation strongly affects the initial eccentricity, particularly in the most central collisions~\cite{Jia:2021tzt,Zhang:2021kxj,Giacalone:2021udy}. Therefore, the final state $v_n$ is expected to be a powerful tool to probe the deformations.

The high order flow coefficients $v_n~(n>3)$ receive contributions not only from the linear response to the initial $\varepsilon_n$ but also from the nonlinear response originated from lower order $\varepsilon_2$ and/or $\varepsilon_3$~\cite{Bhalerao:2014xra,Bhalerao:2013ina,Yan:2015jma}.
For example, the 4$^{\rm th}$ order (complex) anisotropic flow $V_4$ can be decomposed into linear ($V^\mathrm{L}_4$) and nonlinear ($V^\mathrm{NL}_4$) components according to
\begin{equation}
    \begin{split}
    V_4=V^\mathrm{L}_4+V^\mathrm{NL}_4,
    \label{eq:LinearAndNonlinearPart}
    \end{split}
\end{equation}
whose magnitudes are denoted by $v_4^\mathrm{L}$ and $v_{4,22}$, respectively.
The subscript of $v_{4,22}$ represents the part of $v_4$ coming from $\varepsilon_2^2$~\cite{Bhalerao:2014xra,Bhalerao:2013ina,Yan:2015jma}.
In Eq.~(\ref{eq:LinearAndNonlinearPart}) $V^\mathrm{L}_4$ and $V^\mathrm{NL}_4$ are considered to be uncorrelated and $v_{4,22}$ can be measured via a projection of $V_4$ onto the direction of  $V_2$~\cite{Yan:2015jma,ALICE:2017fcd}
\begin{equation}
    \begin{split}
    v_{4,22}&=\frac{\Re\langle V_4(V_2^*)^2\rangle}{\sqrt{\left\langle |V_2|^4\right\rangle}}.
    \label{eq:VnmkInCorrelation}
    \end{split}
\end{equation}
The magnitude of the linear component can be easily derived as $v_4^\mathrm{L} = \sqrt{v_4^2\{2\} - v_{4,22}^2}$.

Furthermore, the correlation between the symmetry planes $\Psi_4$ and $\Psi_2$ can be probed via the nonlinear flow correlation $\rho_{4,22}$ proposed in Ref.~\cite{Yan:2015jma}. It is defined by the ratio of $v_{4,22}$ and $v_4\{2\}$
\begin{equation}
    \rho_{4,22} = \frac{v_{4,22}}{v_4\{2\}}\approx\langle{\cos(4\Psi_4 - 4\Psi_2)}\rangle.
    \label{eq:NonlinearCorrelation}
\end{equation}
In addition, the nonlinear component $V^\mathrm{NL}_4$ can be further decomposed as
\begin{equation}
    \begin{split}
    V^\mathrm{NL}_4&\approx\chi_{4,22}(V_2)^2, \\
    \chi_{4,22}&=\frac{v_{4,22}}{\sqrt{\left\langle |V_2|^4\right\rangle}}=\frac{\Re\langle V_4(V_2^*)^2\rangle}{\left\langle |V_2|^4\right\rangle},
    \label{eq:NonlinearPart}
    \end{split}
\end{equation}
where $\chi_{4,22}$ is called the nonlinear flow-mode coefficient. It represents the strength of nonlinear response to $V_4$ and is independent of $\varepsilon_2$. Recent studies with both transport and hydrodynamic model calculations have shown that nonlinear flow mode observables such as $v_{4,22}$, $\rho_{4,22}$, and $\chi_{4,22}$, owing to their different sensitivities to different stages of heavy-ion collisions~\cite{Bhalerao:2014xra,Zhou:2015eya,Bilandzic:2013kga,Qian:2016fpi,Parkkila:2021tqq,Parkkila:2021yha}, bring distinction power to the study of  deformation of the colliding nuclei~\cite{Zhao:2022uhl,Magdy:2022cvt,Lu:2023fqd}. 

All the observables measured in this study are based on two- and multiparticle correlations, which can be obtained using the {\it Generic~Framework}~\cite{Bilandzic:2013kga,Huo:2017nms,Moravcova:2020wnf} for flow studies.
To suppress non-flow contributions, a pseudorapidity gap \etagap{1.0} was applied in the two-particle correlations in the second harmonic. For high order ($n\geq3$) correlations, a looser pseudorapidity gap of \etagap{0.8} was applied to preserve more particles for the analysis, considering the limited size of the \XeXe data sample.
For the multiparticle correlations, which are less sensitive to non-flow contaminations, \etagap{0.8} was also applied, except for $v_2\{4\}$, where the pseudorapidity gap is unnecessary as their potential non-flow effects are negligible~\cite{ALICE:2014dwt,Moravcova:2020wnf}.

Except $v_2\{2\}$, $v_3\{2\}$, $v_4\{2\}$, and $v_2\{4\}$, which are taken from Ref.~\cite{ALICE:2018lao}, the other observables are measured for the first time in \XeXe collisions. For \PbPb collisions, measurements of most observables were significantly improved after using the entire Run 2 data compared with previous measurements based only on the 2015 data sample~\cite{ALICE:2018rtz,ALICE:2016ccg,ALICE:2020sup,ALICE:2021adw}.

\section{Analysis Details}
\label{sec:AnalysisDetail}

The data sample analysed in this study was recorded by the ALICE detector~\cite{ALICE:2008ngc,ALICE:2004fvi,ALICE:2005vhb,ALICE:2014sbx} during the \XeXe run at \fivefourfour in 2017 and \PbPb runs at \fivenn in 2015 and 2018 at the LHC. Minimum bias events were triggered by the coincidence of two scintillator counter arrays, V0A and V0C~\cite{ALICE:2008ngc,ALICE:2013axi}, covering the pseudorapidity intervals $2.8<\eta<5.1$ and $-3.7<\eta<-1.7$, respectively. Additional \PbPb events in the 0--10\% and 30--50\% centrality classes were recorded in 2018, using central and semicentral triggers, respectively, to maximise the integrated luminosity for central and semiperipheral collisions. 
Pile-up events, where multiple collisions are included in one single event, were rejected using the timing information from the V0 detectors and selections on the correlation of the multiplicity measured by the Inner Tracking System (ITS)~\cite{ALICE:2008ngc, ALICE:2010tia} and the Time Projection Chamber (TPC)~\cite{ALICE:2008ngc,Alme:2010ke}.
Charged particles are reconstructed in the central pseudorapidity region from their hits in the ITS, which is composed of six layers of silicon detectors surrounding the beam vacuum tube, and their energy deposits in the TPC. The track reconstruction in the ITS and the TPC provided the information on the primary vertex. 
The position of the primary vertex along the beam direction, $V_{\rm z}$, was required to be within $\pm$ 10 cm from the centre of the detector. The analysis was performed as a function of collision centrality, determined using the information from the V0 detectors~\cite{ALICE:2013hur,ALICE:2018yvr} and expressed as percentiles of the total inelastic \XeXe or \PbPb cross sections. The whole centrality range considered in this analysis was 0--60\%, where 0\% corresponds to the most central collisions. After the event selection, about 0.8 million \XeXe events and 163 million \PbPb events were analysed in this work.

Charged-particle tracks in the pseudorapidity region $|\eta|<0.8$ and transverse momentum region $0.2<p_{\rm T}<3.0$ GeV/$c$ were selected for the analysis.
The track quality was ensured by requiring at least 70 TPC space points out of a maximum of 159 with an average $\chi^2$ per degree of freedom of the track fit lower than 2.5. The distance of the closest approach (DCA) to the primary vertex in the beam direction, $\mathrm{DCA}_{z}$, was required to be less than 2 cm. In addition, the DCA in the transverse plane was required to be $\mathrm{DCA}_{xy}<0.0105 + 0.0350 p_{\rm T}^{-1.1}$~cm, with $p_{\rm T}$ measured in GeV/$c$, which gives a $p_{\rm T}$-dependent selection on $\mathrm{DCA}_{xy}$ with thresholds at 0.22~cm at 0.2~GeV/$c$ and 0.02~cm at 3.0~GeV/$c$. A $p_{\rm T}$-dependent weight obtained 
from simulations performed with the HIJING event generator~\cite{Wang:1991hta,Gyulassy:1994ew} combined with the GEANT3 transport code~\cite{Brun:1994aa} was applied to correct for the track reconstruction efficiency. The track reconstruction efficiency ranges from 62\% to 80\% for $p_{\rm T}<1.0$ GeV/$c$ and drops slightly for higher $p_{\rm T}$ reaching a roughly constant value of about 76\%. In addition, $\varphi$ distributions of the reconstructed tracks were utilised for extracting a non-uniform acceptance correction.

The sources of systematic uncertainty have been investigated by varying the criteria for selecting events and tracks.
For event selections, the requirement for primary vertex position from the centre of the detector $V_{\rm z}$ was varied to $\pm$5, $\pm$7, and $\pm$9 cm, respectively.
In addition, the centrality estimation was alternatively determined by using the number of hits in the second-most internal layer of the ITS.
In general, these sources yield uncertainties below 1\%; except the uncertainties associated with centrality estimation for $v_{4,22}$, $\rho_{4,22}$, and $\chi_{4,22}$ whose maximum levels reached 1\%.
Furthermore, the systematic effect from pile-up events was studied by varying the selections on the correlations between multiplicities from the ITS and the TPC being found negligible.

Similarly, for the track selections, the minimum number of TPC space points was varied to 80, 90, and 100. The requirement for $\mathrm{DCA}_{xy}$ was changed to $\mathrm{DCA}_{xy}<0.0090 + 0.0300 p_{\rm T}^{-1.1}$~cm, with $p_{\rm T}$ measured in GeV/$c$, while $\mathrm{DCA_{z}}$ was required to be within 1.0 or 0.5~cm. These sources typically result in uncertainties of less than 1\%. Finally, the systematic uncertainties that were statistically significant according to the recommendation in Ref.~\cite{Barlow:2002yb} were added in quadrature to obtain the total systematic uncertainty. The total systematic uncertainties are typically less than 2\% in the 0--60\% centrality range, and they are denoted as grey boxes in the figures in Sec.~\ref{sec:Results}. 

\section{Results}
\label{sec:Results}

\begin{figure}[!htb]
    \begin{center}
      \includegraphics[width=0.8\textwidth]{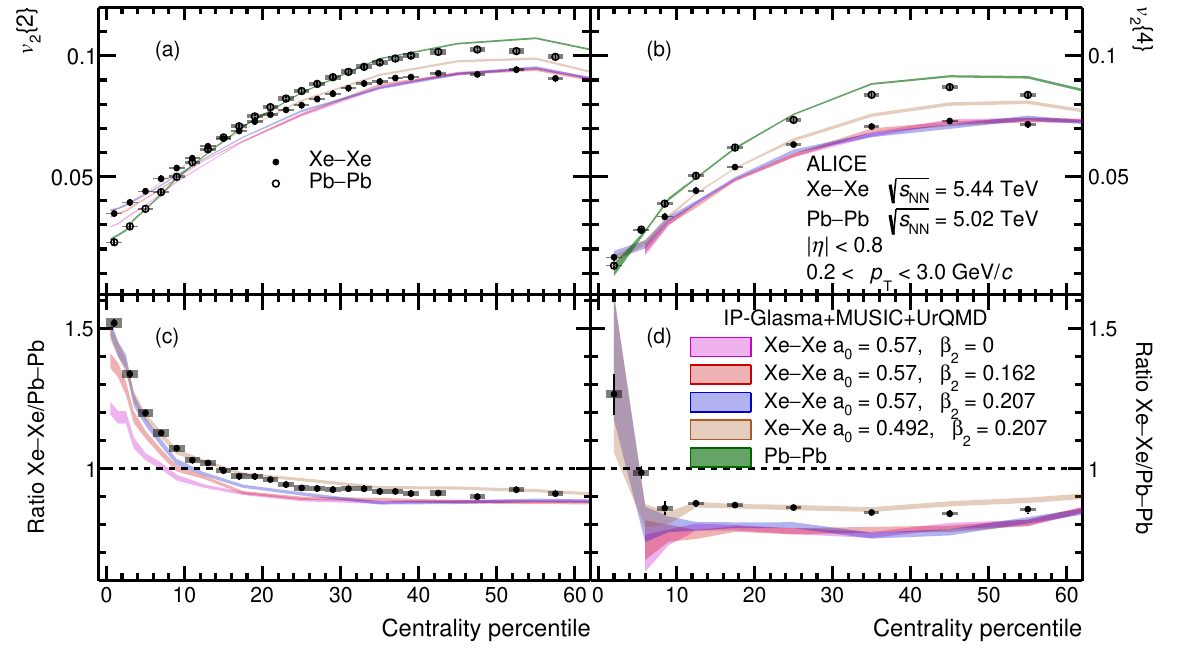}
      \caption[.]{Panels (a) and (b): Charged particle $v_2\{2, \etagap{1.0}\}$ (left) and $v_2\{4\}$ (right) as a function of centrality in \XeXe and \PbPb collisions at \fivefourfour and \fivenn, respectively. Panels (c) and (d): Ratio between \XeXe and \PbPb $v_2\{2, \etagap{1.0}\}$ (left) and $v_2\{4\}$ (right). Statistical and systematical uncertainties are shown as vertical lines and grey boxes, respectively. The measurements are compared with IP-Glasma+MUSIC+UrQMD calculations~\cite{Schenke:2020mbo,Mantysaari:2022ffw} to constrain the $\beta_2$ and $a_0$ parameters of $^{129}$Xe nuclei. The thickness of the bands represent statistical uncertainties.}
      \label{fig:v22v24}
    \end{center}
\end{figure}

Figure~\ref{fig:v22v24} presents the measurements of $v_2\{m\}(m=2,4)$ in \XeXe and \PbPb collisions as a function of centrality. 
In the upper panels, $v_2\{2, \etagap{1.0}\}$ and $v_2\{4\}$ are shown. They increase from central to peripheral \XeXe and \PbPb collisions. The comparisons between Xe--Xe and Pb--Pb results are quantified as ratios in the bottom panels.
Considering the similar dynamic evolution of the created matter in Pb--Pb and Xe--Xe collisions, the ratios of flow observables should largely cancel the final state effects and thus mainly reflect the information on the initial conditions, including the nuclear structure. This has been validated in recent hydrodynamic and transport model calculations~\cite{Lu:2023fqd,Liu:2023pav}.
Both $v_2\{2, \etagap{1.0}\}$ and $v_2\{4\}$ ratios decrease steeply with increasing centrality percentile in central collisions and then level off for midcentral collisions.
The $v_2\{2, \etagap{1.0}\}$ ratio starts at approximately 1.5 in the most central collisions and is larger than unity in the centrality range 0--15\%, whereas the $v_2\{4\}$ ratio starts at approximately 1.3 and is above unity only in the 5\% most central collisions.
In a central collision, the fluctuations of the overlap region play a dominant role, and smaller system size (\XeXe collisions) generates stronger fluctuations~\cite{STAR:2022gki}, which causes both ratios to be larger than unity.
In addition, the deformation of $^{129}$Xe nuclei further enhances $\varepsilon_2$ in ultracentral collisions of 0--5\% centrality; this effect will be discussed in detail later.
In midcentral collisions, $v_2\{2, \etagap{1.0}\}$ and $v_2\{4\}$ ratios remain at approximately 0.9 and 0.85, respectively.
The ratios are below unity due to viscous effects during the medium expansion~\cite{ALICE:2018lao,Molnar:2008xj,Song:2008si}. 

Unlike previous studies~\cite{Bally:2021qys, Dimri:2023wup, Zhao:2024lpc} that investigated nuclear structure based solely on initial-state estimates, the presented measurements are compared with calculations using the sequential combination of the impact-parameter Glasma (IP-Glasma) initial conditions, the MUSIC relativistic hydrodynamic model, and the ultrarelativistic quantum molecular dynamics (UrQMD) model for hadronic rescatterings. This hybrid model is denoted as IP-Glasma+MUSIC+UrQMD~\cite{Schenke:2020mbo, Mantysaari:2022ffw}. 
These calculations are presented as bands of different colours, where the thickness of bands denote the statistical uncertainties of the calculations.
The IP-Glasma+MUSIC+UrQMD model has successfully described particle production and complex anisotropic flow measurements in Pb--Pb collisions at the LHC~\cite{Schenke:2020mbo}, providing valuable insights into both the initial conditions and the dynamical evolution of colliding systems. 
To investigate the impact of nuclear structure, different initial conditions were used for \XeXe calculations, varying the $\beta_2$ quadrupole deformation and the $a_0$ nuclear diffuseness.
The values of $\beta_2$ and $a_0$ were adopted based on existing predictions.
Specifically, $a_0=0.492$ and $\beta_2 =0.207$ are taken from Ref.~\cite{Bally:2021qys}, $\beta_2=0.162$ is from Ref.~\cite{Moller:2015fba}, and $a_0=0.57$ is used in Ref.~\cite{Giacalone:2017dud}. Notably, the setting of $\beta_2=0$ represents a special scenario of a spherical nucleus. Despite the ongoing investigation into the nuclear shape phase transition of $^{129}$Xe, where the $\gamma$-soft structure was discussed~\cite{Zhao:2024lpc}, the current calculations set the $\gamma$ parameter to zero, as all the presented flow observables have been found to be insensitive to the triaxial structure~\cite{Lu:2023fqd}.
For \PbPb calculations, a very weak deformation $\beta_2=0.055$ of $^{\rm 208}$Pb is adopted~\cite{Pritychenko:2013gwa}, which is also used in Ref.~\cite{Bally:2021qys} when the ultra-relativistic energy is considered.
In Fig.~\ref{fig:v22v24}, the IP-Glasma+MUSIC+UrQMD calculations in \PbPb collisions (green bands) align well with the measurements of $v_2\{2, \etagap{1.0}\}$ and $v_2\{4\}$ up to a centrality of 35\%. However, beyond 35\% centrality, the calculated values exceed the measurements.
For \XeXe, in the 0--15\% centrality range, the calculations with $a_0=0.57, \beta_2=0.207$ (blue bands) and $a_0=0.492, \beta_2=0.207$ (brown bands) match the measurements of $v_2\{2, \etagap{1.0}\}$ better, while they underestimate $v_2\{4\}$ in 5--10\% centrality.
Then for the 15--25\% centrality range, the measurements of $v_2\{2, \etagap{1.0}\}$ and $v_2\{4\}$ are better described by the calculations when the parameters are set to $a_0=0.492, \beta_2=0.207$ (brown bands).
Furthermore, in the 35--60\% centrality range, the calculations with $a_0=0.57, \beta_2=0.207$ (blue bands), as well as $a_0=0.57, \beta_2=0.162$ (red bands) and $a_0=0.57, \beta_2=0$ (pink bands) provide better descriptions for the measurements of both $v_2\{2, \etagap{1.0}\}$ and $v_2\{4\}$.
Notably in the 0--10\% centrality range in Fig.~\ref{fig:v22v24}(c), the calculations for $v_2\{2, \etagap{1.0}\}$ with $a_0=0.57, \beta_2 =0.162$ and $a_0=0.57, \beta_2=0$ are approximately 5\% and 20\% lower, respectively, than the measured ratios of \XeXe and \PbPb results. This discrepancy highlights the contributions from the quadrupole deformation of $^{\rm 129}$Xe~\cite{Jia:2021tzt,Zhao:2022uhl,Giacalone:2021udy,Magdy:2022cvt,Lu:2023fqd}.
In this centrality range, the initial shape of the overlapping region is primarily determined by the shape of the colliding nuclei; thus, the deformed nuclei enhance the initial eccentricity $\varepsilon_2$ of the overlapping region, consequently leading to larger $v_2$.

\begin{figure}[!htb]
    \begin{center}
      \includegraphics[width=0.8\textwidth]{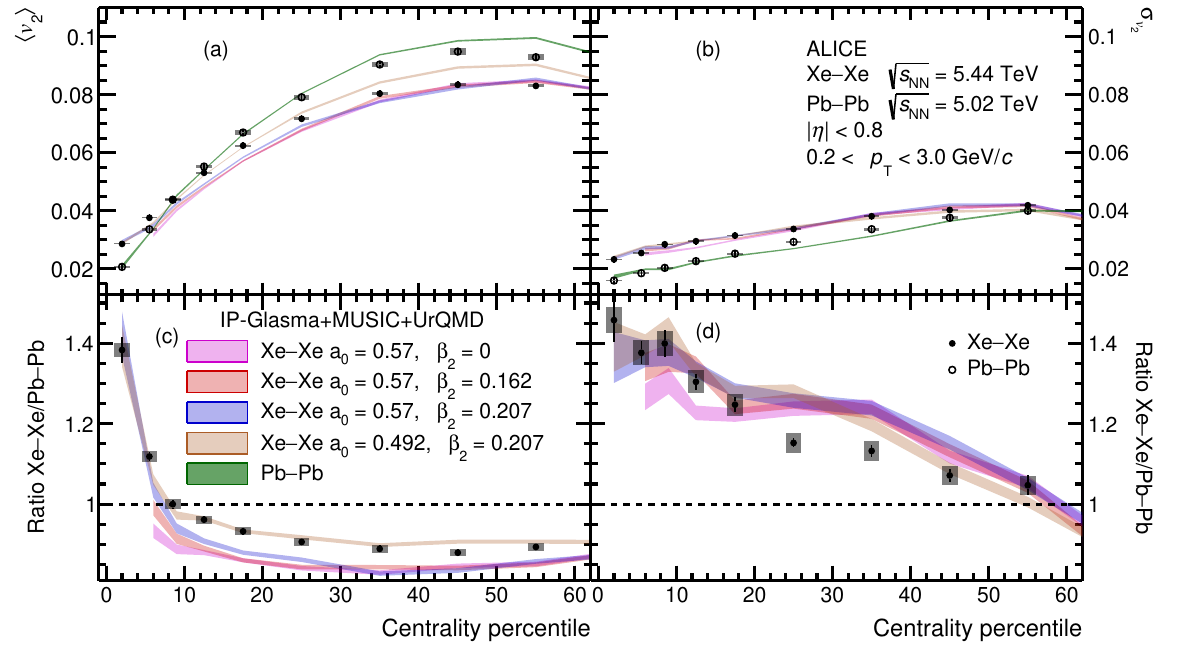}
      \caption[.]{Panels (a) and (b): Charged particle $\langle v_2 \rangle$ (left) and $\sigma_{v_2}$ (right) as a function of centrality in \XeXe and \PbPb collisions at \fivefourfour and \fivenn, respectively. Panels (c) and (d): Ratio between \XeXe and \PbPb $\langle v_2 \rangle$ (left) and $\sigma_{v_2}$ (right). Statistical and systematical uncertainties are shown as vertical lines and grey boxes, respectively. The measurements are compared with IP-Glasma+MUSIC+UrQMD calculations~\cite{Schenke:2020mbo,Mantysaari:2022ffw} to constrain the $\beta_2$ and $a_0$ parameters of $^{129}$Xe nuclei. The thickness of the bands represent statistical uncertainties.}
      \label{fig:MeanSigma}
    \end{center}
\end{figure}
As introduced in Eq.~(\ref{eq:FlowFluctuations}), $v_2\{2\}$ and $v_2\{4\}$ receive contributions from both $\langle v_2 \rangle$ and its event-by-event fluctuations $\sigma_{v_2}$.
Consequently, mean flow and flow fluctuations can be measured separately using the combination of $v_2\{2\}$ and $v_2\{4\}$. 
Figure~\ref{fig:MeanSigma} presents the centrality dependence of $\langle v_2 \rangle$ and $\sigma_{v_2}$ in \XeXe and \PbPb collisions.
In panel (a), $\langle v_2 \rangle$ increases from central to peripheral collisions for both \XeXe and \PbPb collisions. The ratio between \XeXe and \PbPb $\langle v_2 \rangle$ in panel (c) exceeds unity in 0--10\% centrality, then decreases to approximately 0.9 in the midcentral collisions. 
Overall, $\sigma_{v_2}$ in \XeXe is larger than in \PbPb in the 0--60\% centrality range, attributable to the smaller system size of  \XeXe collisions~\cite{STAR:2022gki}.
The ratio between Xe–Xe and Pb–Pb $\sigma_{v_2}$ in panel (d) starts at approximately 1.5 in the most central collisions and steadily decreases with increasing centrality percentile, converging to unity at 60\% centrality.
For $\langle v_2 \rangle$ in Fig.~\ref{fig:MeanSigma}(a) and (c), the IP-Glasma+MUSIC+UrQMD calculations with $\beta_2=0.207$ describe the measurements in 0--10\% centrality.
Due to the extensive statistical samples required, other calculations are only available for centralities above 5\%, which notably underestimate the measured $\langle v_2 \rangle$ for the 0--20\% centrality range.
For $\sigma_{v_2}$ shown in Fig.~\ref{fig:MeanSigma}(b) and (d), most calculations describe the measurements within the presented centrality range, except for the one with $a_0=0.57$ and $\beta_2=0$, which falls below the measurement in 0--20\% centrality. 
A weaker elliptic flow fluctuation $\sigma_{v_2}$ is seen in central Xe--Xe collisions when a spherical nuclear structure of $^{129}$Xe is used in the model calculations.
For centrality above 20\%, the calculations for $\sigma_{v_2}$ with different $a_0$ and $\beta_2$ are compatible with each other within uncertainties, suggesting that $\sigma_{v_2}$ might not depend on the nuclear diffuseness and deformation for non-central collisions.

\begin{figure}[!htb]
    \begin{center}
      \includegraphics[width=0.8\textwidth]{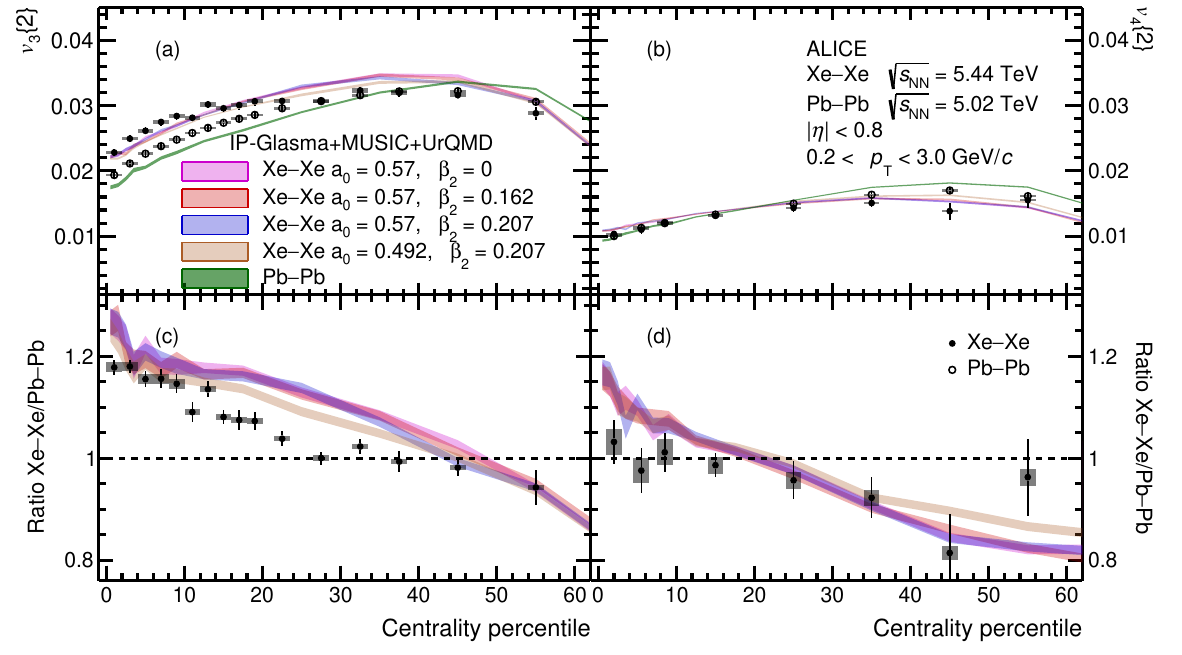}
      \caption[.]{Panels (a) and (b): Charged particle 
$v_3\{2, \etagap{0.8}\}$ (left) and $v_4\{2, \etagap{0.8}\}$ (right) as a function of centrality in \XeXe and \PbPb collisions at \fivefourfour and \fivenn, respectively. Panels (c) and (d): Ratio between \XeXe and \PbPb $v_3\{2, \etagap{0.8}\}$ (left) and $v_4\{2, \etagap{0.8}\}$ (right). Statistical and systematical uncertainties are shown as vertical lines and grey boxes, respectively. The measurements are compared with IP-Glasma+MUSIC+UrQMD calculations~\cite{Schenke:2020mbo,Mantysaari:2022ffw} to constrain the $\beta_2$ and $a_0$ parameters of $^{129}$Xe nuclei. The thickness of the bands represent statistical uncertainties.}
      \label{fig:v3v4}
    \end{center}
\end{figure}
In addition to the study of elliptic flow $v_2$ and its event-by-event fluctuations, the triangular flow $v_3\{2\}$ and quadrangular flow $v_4\{2\}$, which provide more precise constraints on the initial conditions~\cite{Alver:2010dn, Alver:2010gr}, are also examined as a function of centrality in Fig.~\ref{fig:v3v4}. 
In the upper panels, $v_3\{2, \etagap{0.8}\}$ is notably larger in \XeXe than in \PbPb within the 0--35\% centrality range, while the $v_3\{2, \etagap{0.8}\}$ measurements in \XeXe are smaller for more peripheral collisions.
The $v_4\{2, \etagap{0.8}\}$ results are compatible within uncertainties for both \XeXe and \PbPb collisions up to 30\% centrality, after which \XeXe results are smaller than those in \PbPb collisions.
In the lower panels, accordingly, the ratios between \XeXe and \PbPb $v_3\{2, \etagap{0.8}\}$ and $v_4\{2, \etagap{0.8}\}$ decrease steadily with increasing centrality. 
The IP-Glasma+MUSIC+UrQMD calculations are lower than the $v_3\{2, \etagap{0.8}\}$ measurements in \PbPb collisions up to 35\% centrality, beyond which the calculations overestimate the measurements. 
A similar pattern is observed for \XeXe collisions, where the calculations are roughly compatible with the $v_3\{2, \etagap{0.8}\}$ measurements in the central collision and exceed the measured values for centrality above 20\%. 
Meanwhile, no difference is found among the $v_{3}\{2, \etagap{0.8}\}$ calculations with different $\beta_2$ values. 
This is consistent with the expectation that $v_3\{2\}$, which is primarily driven by the linear response to the initial triangularity $\varepsilon_3$~\cite{Niemi:2012aj,Song:2010mg}, may be sensitive to octupole deformation $\beta_3$ but not to quadrupole deformation $\beta_2$. 
This has also been confirmed in the previous AMPT model studies~\cite{Lu:2023fqd}. Furthermore, for the \XeXe/\PbPb ratios in Fig.~\ref{fig:v3v4}, the calculations qualitatively capture the general trend of the centrality dependence of the measured $v_3\{2, \etagap{0.8}\}$ and $v_4\{2, \etagap{0.8}\}$. 
However, all calculations for $v_3\{2, \etagap{0.8}\}$ ratio are higher than the measurements in 10--40\% centrality. 
A distinction is observed between calculations from $a_0 = 0.57$ and $a_0 = 0.492$ in the 10--40\% centrality range; the latter exhibits a slightly better agreement with the measurement. 
Concurrently, the calculations appear to overestimate the $v_4\{2, \etagap{0.8}\}$ ratio in central collisions. 
A difference between the calculations of $v_4\{2, \etagap{0.8}\}$ with $a_0 = 0.57$ and $a_0 = 0.492$ is also noted in more peripheral collisions, as reported from previous AMPT calculations~\cite{Lu:2023fqd, Magdy:2022cvt}. Unfortunately, the significant uncertainties in the measurements preclude a definitive conclusion as to which model calculation better reproduces them. 

\begin{figure}[!htb]
    \begin{center}
      \includegraphics[width=0.8\textwidth]{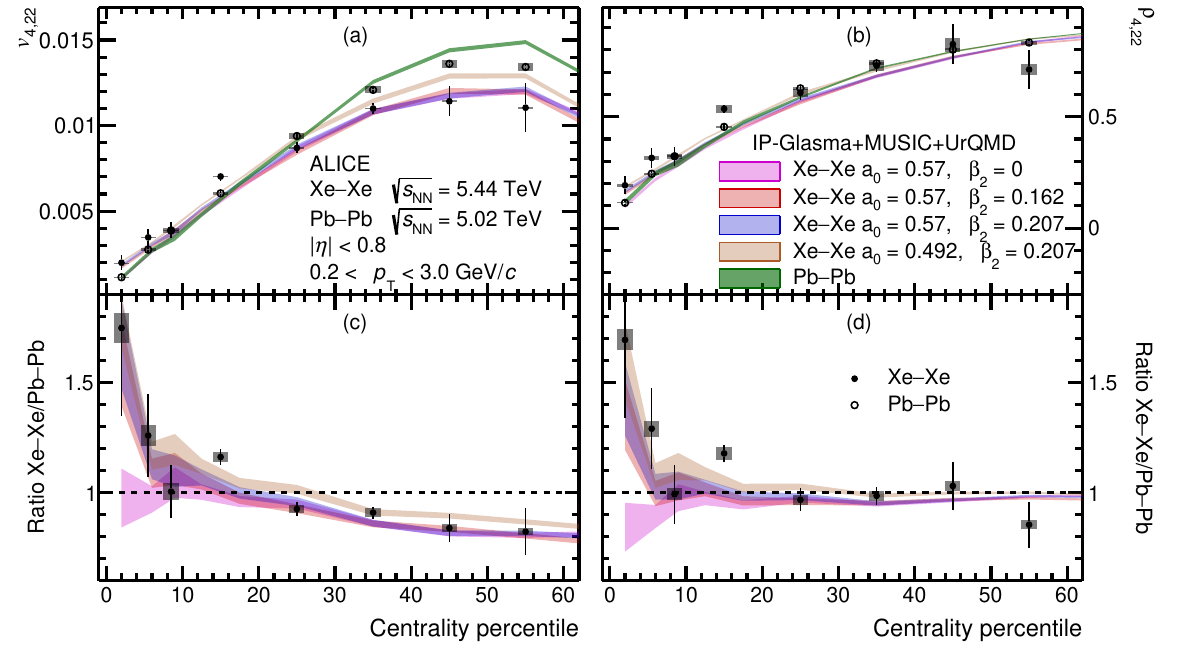}
      \caption[.]{Panels (a) and (b): Charged particle $v_{4,22}$ (left) and $\rho_{4,22}$ (right) as a function of centrality in \XeXe and \PbPb collisions at \fivefourfour and \fivenn, respectively. Panels (c) and (d): Ratio between \XeXe and \PbPb $v_{4,22}$ (left) and $\rho_{4,22}$ (right). Statistical and systematical uncertainties are shown as vertical lines and grey boxes, respectively. The measurements are compared with IP-Glasma+MUSIC+UrQMD calculations~\cite{Schenke:2020mbo,Mantysaari:2022ffw} to constrain the $\beta_2$ and $a_0$ parameters of $^{129}$Xe nuclei. The thickness of the bands represent statistical uncertainties.}
      \label{fig:Harmonic4}
    \end{center}
\end{figure}
Figure~\ref{fig:Harmonic4} shows the centrality dependence of the $v_{4,22}$ nonlinear flow modes \XeXe and \PbPb collisions.
It has been established that $v_{4,22}$ exhibits considerable sensitivities to nuclear deformation parameters~\cite{Lu:2023fqd}, originating from the initial $\varepsilon_{2}^{2}$. 
In the upper panels of Fig.~\ref{fig:Harmonic4}, it can be seen that $v_{4,22}$ increases from central to peripheral \XeXe and \PbPb collisions. 
The $v_{4,22}$ ratio, shown in panel (c) of Fig.~\ref{fig:Harmonic4}, starts at approximately 1.5 in most central collisions and decreases toward more peripheral collisions.
In comparison to the measurements, the IP-Glasma+MUSIC+UrQMD calculations describe $v_{4,22}$ measurements in 0--35\% centrality and only marginally overestimate them in 35--60\% centrality for \PbPb collisions, while they quantitatively capture the $v_{4,22}$ measurements in \XeXe collisions. 
Regarding the ratios in Fig.~\ref{fig:Harmonic4}(c), the measured $v_{4,22}$ ratios in the centrality range 0–20\% are better described by the IP-Glasma+MUSIC+UrQMD calculations with a non-zero $\beta_2$ and are significantly larger than the one with $\beta_2=0$. This aligns with expectations, as $v_{4,22}$ is primarily affected by $\varepsilon_2^2$ in central collisions~\cite{Jia:2021tzt} where $\varepsilon_2$ is influenced mainly by the nuclear quadrupole deformation $\beta_2$.
Additionally, $v_{4,22}$ ratio calculations using $a_0=0.57$ describe the measurements in 20--60\% centrality better, whereas the one with $a_0=0.492$ overestimates the measured $v_{4,22}$ ratio.
A similar observation on the sensitivity of $v_{4,22}$ to $a_0$ in midcentral collisions has been reported in the AMPT studies~\cite{Lu:2023fqd}, suggesting that $v_{4,22}$ serves as a promising probe of the nuclear diffuseness. 

In addition to the nonlinear flow modes, which depend on the magnitudes of $v_{2}$ and/or $v_{3}$, the symmetry plane correlation $\rho_{4,22}$ is investigated in \XeXe and \PbPb collisions. The $\rho_{4,22}$ has been identified as carrying unique sensitivities to the initial conditions of heavy-ion collisions, rendering it a valuable probe for the nuclear structure~\cite{ALICE:2017fcd,ALICE:2020sup}. 
The measurements of $\rho_{4,22}$ are presented as a function of centrality in panels (b) and (d) of Fig.~\ref{fig:Harmonic4}.
In panel (b), $\rho_{4,22}$ shows an increase from central to peripheral collisions in both \XeXe and \PbPb collisions. The $\rho_{4,22}$ ratio drops steeply in the most central collisions, starting from approximately 1.7 down to unity for centralities above 20\%. 
Regarding the ratio of $\rho_{4,22}$ presented in panel (d), the IP-Glasma+MUSIC+UrQMD calculations offer a reasonable description of the measurements, except for the scenario with $\beta_2=0$ in the most central collisions, which assumes a spherical $^{129}$Xe shape and misses the measured $\rho_{4,22}$ ratio. The pronounced correlations between second and fourth-order symmetry planes, $\Psi_2$ and $\Psi_4$, in \XeXe collision, are primarily ascribed to the shape of the colliding nuclei influencing the overlap region in central collisions. A deformed $^{129}$Xe nuclear structure results in an elliptical overlapping region in central collisions, leading to preferred orientations for the symmetry planes rather than random fluctuations, thereby generating stronger correlations between $\Psi_2$ and $\Psi_4$ in \XeXe collisions than in \PbPb collisions. Overall, the IP-Glasma+MUSIC+UrQMD calculations, considering different $a_0$ values, do not exhibit significant differences in $\rho_{4,22}$, taking into account the considerable uncertainties in the model calculations. 

Furthermore, the linear flow mode $v_4^{\rm L}$, the nonlinear flow coefficient $\chi_{4,22}$, and NSC$(3,2)$ have been measured in Xe–Xe collisions at the LHC. These measurements are compared with model calculations of IP-Glasma+MUSIC+UrQMD, which reveal no sensitivity to the variations in nuclear structure. The relevant results are presented in Appendix~\ref{app:SM}.

\begin{figure}[!htb]
    \begin{center}
      \includegraphics[width=0.8\textwidth]{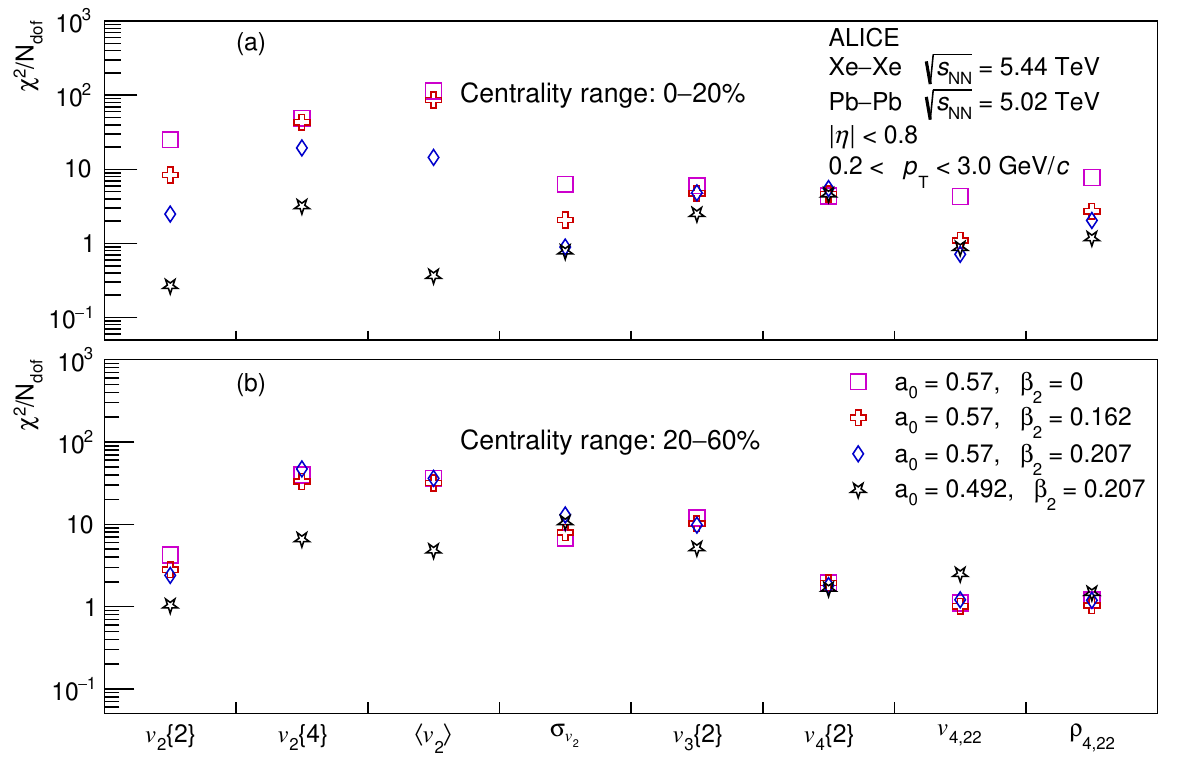}
      \caption[.]{Values of $\chi^2/N_{\rm dof}$ between the measurements (\XeXe/\PbPb) and the calculations (\XeXe/\PbPb). The x-axis represents the different measured observables, and the y-axis is shown on a logarithmic scale. Panels (a) and (b) show the results for the 0--20\% and 20--60\% centrality ranges, respectively.
      }
      \label{fig:chi2ndof}
    \end{center}
\end{figure}
To quantify the agreement between the experimental measurements and the IP-Glasma+MUSIC+UrQMD model calculations with the different configurations, a $\chi^2/N_{\rm dof}$ for each observable was calculated as
\begin{equation}
    \chi^2/N_{\rm dof} = \frac{1}{N_{\rm dof}}\sum \frac{(y_i-f_i)^2}{\sigma_i^2},
    \label{eq:chi2ndof}
\end{equation}
where $y_i$ is the value of the observable experimental measurement at centrality range $i$ and $f_i$ is the value of the observable calculation for the same centrality range with the corresponding configuration, $\sigma_i^2$ is the quadratic sum of the statistical uncertainty $\sigma_{\rm stat}$, systematic uncertainty $\sigma_{\rm sys}$, and model uncertainty $\sigma_{\rm model}$. 
The number of degrees of freedom $N_{\rm dof}$ is obtained by subtracting the number of parameters from the number of data points.
Only the measured ratio (\XeXe/\PbPb) for each observable is considered.
The $\chi^2/N_{\rm dof}$ values for the observables considered in this work are shown in Fig.~\ref{fig:chi2ndof}.
Panel (a) restricts the centrality range to 0--20\%, and panel (b) restricts the centrality range to 20--60\%.
The centrality region is separated because the $\beta_2$ parameter has a strong impact on the observables in central collisions, while the $a_0$ parameter shows influence across the 0--60\% centrality range.
It can be seen that the IP-Glasma+MUSIC+UrQMD calculations with $\beta_2=0.207$ generally provide a better description of the measurements of $v_2$ related observables, as indicated by the smaller $\chi^2/N_{\rm dof}$ values. 
In the 0--20\% centrality range, the calculations with $a_0=0.492, \beta_2=0.207$ yield the smallest $\chi^2/N_{\rm dof}$ for $v_{2}\{2, \etagap{1.0}\}$, $v_{2}\{4\}$ and $\langle v_2 \rangle$, and result in a consistent $\chi^2/N_{\rm dof}$ in comparison to the calculation using $a_0=0.57, \beta_2=0.207$ for $\sigma_{v_2}$. 
This shows the strong influences from $\beta_2$ and $a_0$ on those observables in central collisions.
In the 20--60\% centrality range, the $\chi^2/N_{\rm dof}$ results for $v_2$-related observables are similar for different $\beta_2$ values, indicating that the deformation effect is weak for non-central collisions.
Meanwhile, the calculations with $a_0=0.492, \beta_2=0.207$ still provide the smallest $\chi^2/N_{\rm dof}$ for $v_{2}\{2, \etagap{1.0}\}$, $v_{2}\{4\}$ and $\langle v_2 \rangle$, showing the influences from $a_0$ in midcentral collisions.
In addition, the data-to-model $\chi^2/N_{\rm dof}$ values are shown for the $v_3\{2, \etagap{0.8}\}$ and $v_4$ related observables. The IP-Glasma+MUSIC+UrQMD calculations with $a_0=0.492$ and $\beta_2=0.207$ provide better descriptions of $v_3\{2, \etagap{0.8}\}$, and they also perform reasonably well for $\rho_{4,22}$, compared to the calculations using different $a_0$ or $\beta_2$ parameters. 
In contrast, the calculations with $\beta_2=0$ consistently yield relatively poor descriptions, emphasising the significance of a finite quadrupole deformation for $^{129}$Xe.
For $v_4\{2, \etagap{0.8}\}$, all calculations exhibit similar $\chi^2/N_{\rm dof}$ values, aligning with previous discussions that $v_4\{2, \etagap{0.8}\}$ is not sensitive to the variations in either $a_0$ or $\beta_2$.
For $v_{4,22}$, calculations with $a_0=0.57$ yield smaller $\chi^2/N_{\rm dof}$ values, which are influenced by the large uncertainties in both the model and the measurements. 
Overall, calculations with $a_0=0.492$ and $\beta_2=0.207$ align better with the measurements for the flow observables in \XeXe collisions. 

It is noteworthy that the $\chi^2/N_{\rm dof}$ test might not provide a precise measure but rather qualitatively reflects the potential sensitivities of flow observables to $\beta_2$ and $a_0$. It facilitates the initial exploration of how various flow observables respond to different nuclear structures. Notably, this approach was first applied in complex flow measurements in \PbPb collisions~\cite{ALICE:2020sup}, introducing novel constraints on the tuning of the hydrodynamic framework with varying initial conditions. Subsequently, these flow measurements were incorporated into Global Bayesian fits, leading to the most precise constraints on \PbPb collisions initial conditions to date~\cite{Parkkila:2021tqq}. Therefore, the systematic measurements of complex flow observables presented in this paper are expected to be adopted soon in Bayesian fits, potentially enabling a more reliable extraction of nuclear structure parameters from relativistic nuclear collisions.

\section{Summary}
\label{sec:Summary}

For the first time, measurements of complex flow observables through multiparticle azimuthal correlations have been employed to probe the nuclear structure in heavy-ion collisions. Systematic measurements of various flow observables, including anisotropic flow coefficients ($v_n$), flow fluctuations ($\sigma_{v_2}$), nonlinear and linear components of flow coefficients ($v_{4,22}$, $v_4^{\rm L}$), nonlinear coefficients ($\chi_{4,22}$), correlations between different symmetry planes ($\rho_{4,22}$), and normalised symmetry cumulants have been performed in Xe--Xe and Pb--Pb collisions at \fivefourfour and 5.02 TeV, respectively.
Notably, several flow observables exhibit pronounced differences in the ratio between \XeXe and \PbPb in the most central collisions, which are anticipated from the quadrupole deformation of the $^{129}$Xe nuclear structure. Comprehensive comparisons between the experimental measurements and the IP-Glasma+MUSIC+UrQMD calculations are presented to quantify the effects of quadrupole deformation and nuclear diffuseness. Specifically, the calculations employing different $\beta_2$ quadrupole deformation parameters and $a_0$ nuclear diffuseness parameters are discussed. It has been found that among various IP-Glasma+MUSIC+UrQMD model calculations, the one using $\beta_2=0.207$ generally provides a better description of the flow measurements. Despite noticeable discrepancies between the measurements and the IP-Glasma+MUSIC+UrQMD predictions, the calculations using $a_0=0.492$ seem favoured by the presented measurements. Future Bayesian analysis will allow a more robust extraction of the $\beta_2$ and $a_0$ values. The distinct sensitivities of flow observables to $\beta_2$ and $a_0$ offer valuable insights into constraining the deformation and diffuseness of $^{129}$Xe in its ground state. Systematic measurements of complex flow observables using multiparticle azimuthal correlations at the LHC are opening new avenues for investigating nuclear structure at the energy frontier, complementing low-energy nuclear structure studies and deepening the understanding of fundamental nuclear properties.
Upcoming $^{\rm 16}$O--$^{\rm 16}$O collisions at the LHC will provide novel opportunities to explore the full potential of the LHC on the nuclear structure study probing, in particular, for the first time the $\alpha$-cluster structure of $^{\rm 16}$O at the TeV energy scale~\cite{ALICE:2021wim,Zhang:2024vkh,YuanyuanWang:2024sgp, Zhao:2024feh, Bally:2022vgo}.


\newenvironment{acknowledgement}{\relax}{\relax}
\begin{acknowledgement}
\section*{Acknowledgements}

The ALICE Collaboration would like to thank Chun Shen and Bjoern Schenke for providing the latest calculations from the state-of-the-art models. 
%

The ALICE Collaboration would like to thank all its engineers and technicians for their invaluable contributions to the construction of the experiment and the CERN accelerator teams for the outstanding performance of the LHC complex.
The ALICE Collaboration gratefully acknowledges the resources and support provided by all Grid centres and the Worldwide LHC Computing Grid (WLCG) collaboration.
The ALICE Collaboration acknowledges the following funding agencies for their support in building and running the ALICE detector:
A. I. Alikhanyan National Science Laboratory (Yerevan Physics Institute) Foundation (ANSL), State Committee of Science and World Federation of Scientists (WFS), Armenia;
Austrian Academy of Sciences, Austrian Science Fund (FWF): [M 2467-N36] and Nationalstiftung f\"{u}r Forschung, Technologie und Entwicklung, Austria;
Ministry of Communications and High Technologies, National Nuclear Research Center, Azerbaijan;
Conselho Nacional de Desenvolvimento Cient\'{\i}fico e Tecnol\'{o}gico (CNPq), Financiadora de Estudos e Projetos (Finep), Funda\c{c}\~{a}o de Amparo \`{a} Pesquisa do Estado de S\~{a}o Paulo (FAPESP) and Universidade Federal do Rio Grande do Sul (UFRGS), Brazil;
Bulgarian Ministry of Education and Science, within the National Roadmap for Research Infrastructures 2020-2027 (object CERN), Bulgaria;
Ministry of Education of China (MOEC) , Ministry of Science \& Technology of China (MSTC) and National Natural Science Foundation of China (NSFC), China;
Ministry of Science and Education and Croatian Science Foundation, Croatia;
Centro de Aplicaciones Tecnol\'{o}gicas y Desarrollo Nuclear (CEADEN), Cubaenerg\'{\i}a, Cuba;
Ministry of Education, Youth and Sports of the Czech Republic, Czech Republic;
The Danish Council for Independent Research | Natural Sciences, the VILLUM FONDEN and Danish National Research Foundation (DNRF), Denmark;
Helsinki Institute of Physics (HIP), Finland;
Commissariat \`{a} l'Energie Atomique (CEA) and Institut National de Physique Nucl\'{e}aire et de Physique des Particules (IN2P3) and Centre National de la Recherche Scientifique (CNRS), France;
Bundesministerium f\"{u}r Bildung und Forschung (BMBF) and GSI Helmholtzzentrum f\"{u}r Schwerionenforschung GmbH, Germany;
General Secretariat for Research and Technology, Ministry of Education, Research and Religions, Greece;
National Research, Development and Innovation Office, Hungary;
Department of Atomic Energy Government of India (DAE), Department of Science and Technology, Government of India (DST), University Grants Commission, Government of India (UGC) and Council of Scientific and Industrial Research (CSIR), India;
National Research and Innovation Agency - BRIN, Indonesia;
Istituto Nazionale di Fisica Nucleare (INFN), Italy;
Japanese Ministry of Education, Culture, Sports, Science and Technology (MEXT) and Japan Society for the Promotion of Science (JSPS) KAKENHI, Japan;
Consejo Nacional de Ciencia (CONACYT) y Tecnolog\'{i}a, through Fondo de Cooperaci\'{o}n Internacional en Ciencia y Tecnolog\'{i}a (FONCICYT) and Direcci\'{o}n General de Asuntos del Personal Academico (DGAPA), Mexico;
Nederlandse Organisatie voor Wetenschappelijk Onderzoek (NWO), Netherlands;
The Research Council of Norway, Norway;
Pontificia Universidad Cat\'{o}lica del Per\'{u}, Peru;
Ministry of Science and Higher Education, National Science Centre and WUT ID-UB, Poland;
Korea Institute of Science and Technology Information and National Research Foundation of Korea (NRF), Republic of Korea;
Ministry of Education and Scientific Research, Institute of Atomic Physics, Ministry of Research and Innovation and Institute of Atomic Physics and Universitatea Nationala de Stiinta si Tehnologie Politehnica Bucuresti, Romania;
Ministry of Education, Science, Research and Sport of the Slovak Republic, Slovakia;
National Research Foundation of South Africa, South Africa;
Swedish Research Council (VR) and Knut \& Alice Wallenberg Foundation (KAW), Sweden;
European Organization for Nuclear Research, Switzerland;
Suranaree University of Technology (SUT), National Science and Technology Development Agency (NSTDA) and National Science, Research and Innovation Fund (NSRF via PMU-B B05F650021), Thailand;
Turkish Energy, Nuclear and Mineral Research Agency (TENMAK), Turkey;
National Academy of  Sciences of Ukraine, Ukraine;
Science and Technology Facilities Council (STFC), United Kingdom;
National Science Foundation of the United States of America (NSF) and United States Department of Energy, Office of Nuclear Physics (DOE NP), United States of America.
In addition, individual groups or members have received support from:
Czech Science Foundation (grant no. 23-07499S), Czech Republic;
FORTE project, reg.\ no.\ CZ.02.01.01/00/22\_008/0004632, Czech Republic, co-funded by the European Union, Czech Republic;
European Research Council (grant no. 950692), European Union;
ICSC - Centro Nazionale di Ricerca in High Performance Computing, Big Data and Quantum Computing, European Union - NextGenerationEU;
Academy of Finland (Center of Excellence in Quark Matter) (grant nos. 346327, 346328), Finland;
Deutsche Forschungs Gemeinschaft (DFG, German Research Foundation) ``Neutrinos and Dark Matter in Astro- and Particle Physics'' (grant no. SFB 1258), Germany.

\end{acknowledgement}

\bibliographystyle{utphys}   
\bibliography{bibliography}

\providecommand{\href}[2]{#2}\begingroup\raggedright\begin{thebibliography}{100}

\bibitem{Lu:2013ena}
Z.~T. Lu, P.~Mueller, G.~W.~F. Drake, W.~Noertershaeuser, S.~C. Pieper, and
  Z.~C. Yan, ``{Colloquium: Laser probing of neutron-rich nuclei in light
  atoms}'', \href{https://doi.org/10.1103/RevModPhys.85.1383}{{\em Rev. Mod.
  Phys.} {\bfseries 85} (2013) 1383--1400},
  \href{https://arxiv.org/abs/1307.2872}{{\ttfamily arXiv:1307.2872
  [nucl-ex]}}.

\bibitem{Heylen:2020cco}
H.~Heylen {\em et~al.}, ``{High-resolution laser spectroscopy of
  $^{27-32}$Al}'', \href{https://doi.org/10.1103/PhysRevC.103.014318}{{\em
  Phys. Rev. C} {\bfseries 103} (2021) 014318},
  \href{https://arxiv.org/abs/2010.06918}{{\ttfamily arXiv:2010.06918
  [nucl-ex]}}.

\bibitem{Bree:2014mxa}
N.~Bree {\em et~al.}, ``{Shape Coexistence in the Neutron-Deficient Even-Even
  $^{182-188}$Hg Isotopes Studied via Coulomb Excitation}'',
  \href{https://doi.org/10.1103/PhysRevLett.112.162701}{{\em Phys. Rev. Lett.}
  {\bfseries 112} (2014) 162701}.

\bibitem{Ayangeakaa:2019psv}
A.~D. Ayangeakaa {\em et~al.}, ``{Evidence for Rigid Triaxial Deformation in
  $^{76}$Ge from a Model-Independent Analysis}'',
  \href{https://doi.org/10.1103/PhysRevLett.123.102501}{{\em Phys. Rev. Lett.}
  {\bfseries 123} (2019) 102501},
  \href{https://arxiv.org/abs/1909.03270}{{\ttfamily arXiv:1909.03270
  [nucl-ex]}}.

\bibitem{Koszorus:2020mgn}
A.~Koszor\'us {\em et~al.}, ``{Charge radii of exotic potassium isotopes
  challenge nuclear theory and the magic character of $N = 32$}'',
  \href{https://doi.org/10.1038/s41567-020-01136-5}{{\em Nature Phys.}
  {\bfseries 17} (2021) 439--443},
  \href{https://arxiv.org/abs/2012.01864}{{\ttfamily arXiv:2012.01864
  [nucl-ex]}}. [Erratum: Nature Phys. 17, 539 (2021)].

\bibitem{Warbinek:2024ncq}
J.~Warbinek {\em et~al.}, ``{Smooth trends in fermium charge radii and the
  impact of shell effects}'',
  \href{https://doi.org/10.1038/s41586-024-08062-z}{{\em Nature} {\bfseries
  634} (2024) 1075--1079}.

\bibitem{Hergert:2020bxy}
H.~Hergert, ``{A Guided Tour of $ab$ $initio$ Nuclear Many-Body Theory}'',
  \href{https://doi.org/10.3389/fphy.2020.00379}{{\em Front. in Phys.}
  {\bfseries 8} (2020) 379}, \href{https://arxiv.org/abs/2008.05061}{{\ttfamily
  arXiv:2008.05061 [nucl-th]}}.

\bibitem{Gandolfi:2020pbj}
S.~Gandolfi, D.~Lonardoni, A.~Lovato, and M.~Piarulli, ``{Atomic nuclei from
  quantum Monte Carlo calculations with chiral EFT interactions}'',
  \href{https://doi.org/10.3389/fphy.2020.00117}{{\em Front. in Phys.}
  {\bfseries 8} (2020) 117}, \href{https://arxiv.org/abs/2001.01374}{{\ttfamily
  arXiv:2001.01374 [nucl-th]}}.

\bibitem{Soma:2020xhv}
V.~Som\`a, ``{Self-consistent Green's function theory for atomic nuclei}'',
  \href{https://doi.org/10.3389/fphy.2020.00340}{{\em Front. in Phys.}
  {\bfseries 8} (2020) 340}, \href{https://arxiv.org/abs/2003.11321}{{\ttfamily
  arXiv:2003.11321 [nucl-th]}}.

\bibitem{Lahde:2019npb}
T.~A. L\"ahde and U.-G. Mei\ss{}ner,
  \href{https://doi.org/10.1007/978-3-030-14189-9}{{\em {Nuclear Lattice
  Effective Field Theory: An introduction}}}, vol.~957.
\newblock Springer, 2019.

\bibitem{Ekstrom:2022yea}
A.~Ekstr\"om, C.~Forss\'en, G.~Hagen, G.~R. Jansen, W.~Jiang, and
  T.~Papenbrock, ``{What is ab initio in nuclear theory?}'',
  \href{https://doi.org/10.3389/fphy.2023.1129094}{{\em Front. Phys.}
  {\bfseries 11} (2023) 1129094},
  \href{https://arxiv.org/abs/2212.11064}{{\ttfamily arXiv:2212.11064
  [nucl-th]}}.

\bibitem{Hu:2021trw}
B.~Hu {\em et~al.}, ``{Ab initio predictions link the neutron skin of
  $^{208}$Pb to nuclear forces}'',
  \href{https://doi.org/10.1038/s41567-023-02324-9}{{\em Nature Phys.}
  {\bfseries 18} (2022) 1196--1200},
  \href{https://arxiv.org/abs/2112.01125}{{\ttfamily arXiv:2112.01125
  [nucl-th]}}.

\bibitem{Miyagi:2021pdc}
T.~Miyagi, S.~R. Stroberg, P.~Navr\'atil, K.~Hebeler, and J.~D. Holt,
  ``{Converged ab initio calculations of heavy nuclei}'',
  \href{https://doi.org/10.1103/PhysRevC.105.014302}{{\em Phys. Rev. C}
  {\bfseries 105} (2022) 014302},
  \href{https://arxiv.org/abs/2104.04688}{{\ttfamily arXiv:2104.04688
  [nucl-th]}}.

\bibitem{STAR:2015mki}
{\bfseries STAR} Collaboration, L.~Adamczyk {\em et~al.}, ``{Azimuthal
  anisotropy in U$+$U and Au$+$Au collisions at RHIC}'',
  \href{https://doi.org/10.1103/PhysRevLett.115.222301}{{\em Phys. Rev. Lett.}
  {\bfseries 115} (2015) 222301},
  \href{https://arxiv.org/abs/1505.07812}{{\ttfamily arXiv:1505.07812
  [nucl-ex]}}.

\bibitem{STAR:2021mii}
{\bfseries STAR} Collaboration, M.~Abdallah {\em et~al.}, ``{Search for the
  chiral magnetic effect with isobar collisions at $\sqrt {s_{\rm NN}}$=200 GeV
  by the STAR Collaboration at the BNL Relativistic Heavy Ion Collider}'',
  \href{https://doi.org/10.1103/PhysRevC.105.014901}{{\em Phys. Rev. C}
  {\bfseries 105} (2022) 014901},
  \href{https://arxiv.org/abs/2109.00131}{{\ttfamily arXiv:2109.00131
  [nucl-ex]}}.

\bibitem{Zhang:2021kxj}
C.~Zhang and J.~Jia, ``{Evidence of Quadrupole and Octupole Deformations in
  $^{96}$Zr$+$$^{96}$Zr and $^{96}$Ru$+$$^{96}$Ru Collisions at
  Ultrarelativistic Energies}'',
  \href{https://doi.org/10.1103/PhysRevLett.128.022301}{{\em Phys. Rev. Lett.}
  {\bfseries 128} (2022) 022301},
  \href{https://arxiv.org/abs/2109.01631}{{\ttfamily arXiv:2109.01631
  [nucl-th]}}.

\bibitem{STAR:2024wgy}
{\bfseries STAR} Collaboration, M.~I. Abdulhamid {\em et~al.}, ``{Imaging
  shapes of atomic nuclei in high-energy nuclear collisions}'',
  \href{https://doi.org/10.1038/s41586-024-08097-2}{{\em Nature} {\bfseries
  635} (2024) 67--72}, \href{https://arxiv.org/abs/2401.06625}{{\ttfamily
  arXiv:2401.06625 [nucl-ex]}}.

\bibitem{Zhao:2024feh}
X.-L. Zhao, G.-L. Ma, Y.~Zhou, Z.-W. Lin, and C.~Zhang, ``{Nuclear cluster
  structure effect in $^{16}$O+$^{16}$O collisions at the top RHIC energy}'',
  \href{https://arxiv.org/abs/2404.09780}{{\ttfamily arXiv:2404.09780
  [nucl-th]}}.

\bibitem{ALICE:2018lao}
{\bfseries ALICE} Collaboration, S.~Acharya {\em et~al.}, ``{Anisotropic flow
  in Xe-Xe collisions at $\sqrt{s_{_{\rm NN}}} = 5.44$ TeV}'',
  \href{https://doi.org/10.1016/j.physletb.2018.06.059}{{\em Phys. Lett. B}
  {\bfseries 784} (2018) 82--95},
  \href{https://arxiv.org/abs/1805.01832}{{\ttfamily arXiv:1805.01832
  [nucl-ex]}}.

\bibitem{ALICE:2018yvr}
{\bfseries ALICE} Collaboration, S.~Acharya {\em et~al.}, ``{Centrality
  determination using the Glauber model in Xe-Xe collisions at $\sqrt{s_{_{\rm
  NN}}} = 5.44$ TeV}'',
  \href{https://arxiv.org/abs/ALICE-PUBLIC-2018-003}{{\ttfamily
  ALICE-PUBLIC-2018-003}}. \url{https://cds.cern.ch/record/2315401}.

\bibitem{ALICE:2018cpu}
{\bfseries ALICE} Collaboration, S.~Acharya {\em et~al.}, ``{Centrality and
  pseudorapidity dependence of the charged-particle multiplicity density in
  Xe\textendash{}Xe collisions at $\sqrt{s_{_{\rm NN}}} = 5.44$ TeV}'',
  \href{https://doi.org/10.1016/j.physletb.2018.12.048}{{\em Phys. Lett. B}
  {\bfseries 790} (2019) 35--48},
  \href{https://arxiv.org/abs/1805.04432}{{\ttfamily arXiv:1805.04432
  [nucl-ex]}}.

\bibitem{ATLAS:2022dov}
{\bfseries ATLAS} Collaboration, G.~Aad {\em et~al.}, ``{Correlations between
  flow and transverse momentum in Xe$+$Xe and Pb+Pb collisions at the LHC with
  the ATLAS detector: A probe of the heavy-ion initial state and nuclear
  deformation}'', \href{https://doi.org/10.1103/PhysRevC.107.054910}{{\em Phys.
  Rev. C} {\bfseries 107} (2023) 054910},
  \href{https://arxiv.org/abs/2205.00039}{{\ttfamily arXiv:2205.00039
  [nucl-ex]}}.

\bibitem{CMS:2019cyz}
{\bfseries CMS} Collaboration, A.~M. Sirunyan {\em et~al.}, ``{Charged-particle
  angular correlations in XeXe collisions at $\sqrt{s_{_{\rm NN}}} = 5.44$
  TeV}'', \href{https://doi.org/10.1103/PhysRevC.100.044902}{{\em Phys. Rev. C}
  {\bfseries 100} (2019) 044902},
  \href{https://arxiv.org/abs/1901.07997}{{\ttfamily arXiv:1901.07997
  [hep-ex]}}.

\bibitem{Giacalone:2021udy}
G.~Giacalone, J.~Jia, and C.~Zhang, ``{Impact of Nuclear Deformation on
  Relativistic Heavy-Ion Collisions: Assessing Consistency in Nuclear Physics
  across Energy Scales}'',
  \href{https://doi.org/10.1103/PhysRevLett.127.242301}{{\em Phys. Rev. Lett.}
  {\bfseries 127} (2021) 242301},
  \href{https://arxiv.org/abs/2105.01638}{{\ttfamily arXiv:2105.01638
  [nucl-th]}}.

\bibitem{Zhao:2022uhl}
S.~Zhao, H.-j. Xu, Y.-X. Liu, and H.~Song, ``{Probing the nuclear deformation
  with three-particle asymmetric cumulant in RHIC isobar runs}'',
  \href{https://doi.org/10.1016/j.physletb.2023.137838}{{\em Phys. Lett. B}
  {\bfseries 839} (2023) 137838},
  \href{https://arxiv.org/abs/2204.02387}{{\ttfamily arXiv:2204.02387
  [nucl-th]}}.

\bibitem{ALICE:2021gxt}
{\bfseries ALICE} Collaboration, S.~Acharya {\em et~al.}, ``{Characterizing the
  initial conditions of heavy-ion collisions at the LHC with mean transverse
  momentum and anisotropic flow correlations}'',
  \href{https://doi.org/10.1016/j.physletb.2022.137393}{{\em Phys. Lett. B}
  {\bfseries 834} (2022) 137393},
  \href{https://arxiv.org/abs/2111.06106}{{\ttfamily arXiv:2111.06106
  [nucl-ex]}}.

\bibitem{Bally:2021qys}
B.~Bally, M.~Bender, G.~Giacalone, and V.~Som\`a, ``{Evidence of the triaxial
  structure of $\boldsymbol{^{129}}$Xe at the Large Hadron Collider}'',
  \href{https://doi.org/10.1103/PhysRevLett.128.082301}{{\em Phys. Rev. Lett.}
  {\bfseries 128} (2022) 082301},
  \href{https://arxiv.org/abs/2108.09578}{{\ttfamily arXiv:2108.09578
  [nucl-th]}}.

\bibitem{Xu:2024bdh}
H.-j. Xu, J.~Zhao, and F.~Wang, ``{Hexadecapole Deformation of U238 from
  Relativistic Heavy-Ion Collisions Using a Nonlinear Response Coefficient}'',
  \href{https://doi.org/10.1103/PhysRevLett.132.262301}{{\em Phys. Rev. Lett.}
  {\bfseries 132} (2024) 262301},
  \href{https://arxiv.org/abs/2402.16550}{{\ttfamily arXiv:2402.16550
  [nucl-th]}}.

\bibitem{Ryssens:2023fkv}
W.~Ryssens, G.~Giacalone, B.~Schenke, and C.~Shen, ``{Evidence of Hexadecapole
  Deformation in Uranium-238 at the Relativistic Heavy Ion Collider}'',
  \href{https://doi.org/10.1103/PhysRevLett.130.212302}{{\em Phys. Rev. Lett.}
  {\bfseries 130} (2023) 212302},
  \href{https://arxiv.org/abs/2302.13617}{{\ttfamily arXiv:2302.13617
  [nucl-th]}}.

\bibitem{Giacalone:2023cet}
G.~Giacalone, G.~Nijs, and W.~van~der Schee, ``{Determination of the Neutron
  Skin of $^{208}$Pb from Ultrarelativistic Nuclear Collisions}'',
  \href{https://doi.org/10.1103/PhysRevLett.131.202302}{{\em Phys. Rev. Lett.}
  {\bfseries 131} (2023) 202302},
  \href{https://arxiv.org/abs/2305.00015}{{\ttfamily arXiv:2305.00015
  [nucl-th]}}.

\bibitem{Li:2019kkh}
H.~Li, H.-j. Xu, Y.~Zhou, X.~Wang, J.~Zhao, L.-W. Chen, and F.~Wang, ``{Probing
  the neutron skin with ultrarelativistic isobaric collisions}'',
  \href{https://doi.org/10.1103/PhysRevLett.125.222301}{{\em Phys. Rev. Lett.}
  {\bfseries 125} (2020) 222301},
  \href{https://arxiv.org/abs/1910.06170}{{\ttfamily arXiv:1910.06170
  [nucl-th]}}.

\bibitem{Xu:2021uar}
H.-j. Xu, W.~Zhao, H.~Li, Y.~Zhou, L.-W. Chen, and F.~Wang, ``{Probing nuclear
  structure with mean transverse momentum in relativistic isobar collisions}'',
  \href{https://doi.org/10.1103/PhysRevC.108.L011902}{{\em Phys. Rev. C}
  {\bfseries 108} (2023) L011902},
  \href{https://arxiv.org/abs/2111.14812}{{\ttfamily arXiv:2111.14812
  [nucl-th]}}.

\bibitem{Jia:2021qyu}
J.~Jia, ``{Probing triaxial deformation of atomic nuclei in high-energy heavy
  ion collisions}'', \href{https://doi.org/10.1103/PhysRevC.105.044905}{{\em
  Phys. Rev. C} {\bfseries 105} (2022) 044905},
  \href{https://arxiv.org/abs/2109.00604}{{\ttfamily arXiv:2109.00604
  [nucl-th]}}.

\bibitem{Jia:2021tzt}
J.~Jia, ``{Shape of atomic nuclei in heavy ion collisions}'',
  \href{https://doi.org/10.1103/PhysRevC.105.014905}{{\em Phys. Rev. C}
  {\bfseries 105} (2022) 014905},
  \href{https://arxiv.org/abs/2106.08768}{{\ttfamily arXiv:2106.08768
  [nucl-th]}}.

\bibitem{Magdy:2022cvt}
N.~Magdy, ``{Impact of nuclear deformation on collective flow observables in
  relativistic U+U collisions}'',
  \href{https://doi.org/10.1140/epja/s10050-023-00982-0}{{\em Eur. Phys. J. A}
  {\bfseries 59} (2023) 64}, \href{https://arxiv.org/abs/2206.05332}{{\ttfamily
  arXiv:2206.05332 [nucl-th]}}.

\bibitem{Nielsen:2023znu}
E.~G.~D. Nielsen, F.~K. R\o{}mer, K.~Gulbrandsen, and Y.~Zhou, ``{Generic
  multi-particle transverse momentum correlations as a new tool for studying
  nuclear structure at the energy frontier}'',
  \href{https://doi.org/10.1140/epja/s10050-024-01266-x}{{\em Eur. Phys. J. A}
  {\bfseries 60} (2024) 38}, \href{https://arxiv.org/abs/2312.00492}{{\ttfamily
  arXiv:2312.00492 [nucl-th]}}.

\bibitem{Lu:2023fqd}
Z.~Lu, M.~Zhao, X.~Li, J.~Jia, and Y.~Zhou, ``{Probe nuclear structure using
  the anisotropic flow at the Large Hadron Collider}'',
  \href{https://doi.org/10.1140/epja/s10050-023-01194-2}{{\em Eur. Phys. J. A}
  {\bfseries 59} (2023) 279},
  \href{https://arxiv.org/abs/2309.09663}{{\ttfamily arXiv:2309.09663
  [nucl-th]}}.

\bibitem{Zhao:2024lpc}
S.~Zhao, H.-j. Xu, Y.~Zhou, Y.-X. Liu, and H.~Song, ``{Exploring the Nuclear
  Shape Phase Transition in Ultra-Relativistic $^{129}$Xe+$^{129}$Xe Collisions
  at the LHC}'', \href{https://arxiv.org/abs/2403.07441}{{\ttfamily
  arXiv:2403.07441 [nucl-th]}}.

\bibitem{Muller:2012zq}
B.~Muller, J.~Schukraft, and B.~Wyslouch, ``{First Results from Pb+Pb
  collisions at the LHC}'',
  \href{https://doi.org/10.1146/annurev-nucl-102711-094910}{{\em Ann. Rev.
  Nucl. Part. Sci.} {\bfseries 62} (2012) 361--386},
  \href{https://arxiv.org/abs/1202.3233}{{\ttfamily arXiv:1202.3233 [hep-ex]}}.

\bibitem{Drescher:2007cd}
H.-J. Drescher, A.~Dumitru, C.~Gombeaud, and J.-Y. Ollitrault, ``{The
  Centrality dependence of elliptic flow, the hydrodynamic limit, and the
  viscosity of hot QCD}'',
  \href{https://doi.org/10.1103/PhysRevC.76.024905}{{\em Phys. Rev. C}
  {\bfseries 76} (2007) 024905},
  \href{https://arxiv.org/abs/0704.3553}{{\ttfamily arXiv:0704.3553
  [nucl-th]}}.

\bibitem{Heinz:2013th}
U.~Heinz and R.~Snellings, ``{Collective flow and viscosity in relativistic
  heavy-ion collisions}'',
  \href{https://doi.org/10.1146/annurev-nucl-102212-170540}{{\em Ann. Rev.
  Nucl. Part. Sci.} {\bfseries 63} (2013) 123--151},
  \href{https://arxiv.org/abs/1301.2826}{{\ttfamily arXiv:1301.2826
  [nucl-th]}}.

\bibitem{Molnar:2001ux}
D.~Molnar and M.~Gyulassy, ``{Saturation of elliptic flow and the transport
  opacity of the gluon plasma at RHIC}'',
  \href{https://doi.org/10.1016/S0375-9474(01)01224-6}{{\em Nucl. Phys. A}
  {\bfseries 697} (2002) 495--520},
  \href{https://arxiv.org/abs/nucl-th/0104073}{{\ttfamily
  arXiv:nucl-th/0104073}}. [Erratum: Nucl.Phys.A 703, 893--894 (2002)].

\bibitem{Song:2017wtw}
H.~Song, Y.~Zhou, and K.~Gajdosova, ``{Collective flow and hydrodynamics in
  large and small systems at the LHC}'',
  \href{https://doi.org/10.1007/s41365-017-0245-4}{{\em Nucl. Sci. Tech.}
  {\bfseries 28} (2017) 99}, \href{https://arxiv.org/abs/1703.00670}{{\ttfamily
  arXiv:1703.00670 [nucl-th]}}.

\bibitem{ALICE:2022wpn}
{\bfseries ALICE} Collaboration, S.~Acharya {\em et~al.}, ``{The ALICE
  experiment: a journey through QCD}'',
  \href{https://doi.org/10.1140/epjc/s10052-024-12935-y}{{\em Eur. Phys. J. C}
  {\bfseries 84} (2024) 813},
  \href{https://arxiv.org/abs/2211.04384}{{\ttfamily arXiv:2211.04384
  [nucl-ex]}}.

\bibitem{Voloshin:1994mz}
S.~Voloshin and Y.~Zhang, ``{Flow study in relativistic nuclear collisions by
  Fourier expansion of Azimuthal particle distributions}'',
  \href{https://doi.org/10.1007/s002880050141}{{\em Z. Phys. C} {\bfseries 70}
  (1996) 665--672}, \href{https://arxiv.org/abs/hep-ph/9407282}{{\ttfamily
  arXiv:hep-ph/9407282}}.

\bibitem{ALICE:2011ab}
{\bfseries ALICE} Collaboration, K.~Aamodt {\em et~al.}, ``{Higher harmonic
  anisotropic flow measurements of charged particles in Pb--Pb collisions at
  $\sqrt{s_{_{\rm NN}}} = 2.76$ TeV}'',
  \href{https://doi.org/10.1103/PhysRevLett.107.032301}{{\em Phys. Rev. Lett.}
  {\bfseries 107} (2011) 032301},
  \href{https://arxiv.org/abs/1105.3865}{{\ttfamily arXiv:1105.3865
  [nucl-ex]}}.

\bibitem{ALICE:2016ccg}
{\bfseries ALICE} Collaboration, J.~Adam {\em et~al.}, ``{Anisotropic flow of
  charged particles in Pb--Pb collisions at $\sqrt{s_{_{\rm NN}}}=5.02$ TeV}'',
  \href{https://doi.org/10.1103/PhysRevLett.116.132302}{{\em Phys. Rev. Lett.}
  {\bfseries 116} (2016) 132302},
  \href{https://arxiv.org/abs/1602.01119}{{\ttfamily arXiv:1602.01119
  [nucl-ex]}}.

\bibitem{ALICE:2019zfl}
{\bfseries ALICE} Collaboration, S.~Acharya {\em et~al.}, ``{Investigations of
  Anisotropic Flow Using Multiparticle Azimuthal Correlations in pp, p--Pb,
  Xe--Xe, and Pb--Pb Collisions at the LHC}'',
  \href{https://doi.org/10.1103/PhysRevLett.123.142301}{{\em Phys. Rev. Lett.}
  {\bfseries 123} (2019) 142301},
  \href{https://arxiv.org/abs/1903.01790}{{\ttfamily arXiv:1903.01790
  [nucl-ex]}}.

\bibitem{ATLAS:2012at}
{\bfseries ATLAS} Collaboration, G.~Aad {\em et~al.}, ``{Measurement of the
  azimuthal anisotropy for charged particle production in $\sqrt{s_{_{\rm
  NN}}}=2.76$ TeV lead-lead collisions with the ATLAS detector}'',
  \href{https://doi.org/10.1103/PhysRevC.86.014907}{{\em Phys. Rev. C}
  {\bfseries 86} (2012) 014907},
  \href{https://arxiv.org/abs/1203.3087}{{\ttfamily arXiv:1203.3087 [hep-ex]}}.

\bibitem{ATLAS:2019dct}
{\bfseries ATLAS} Collaboration, G.~Aad {\em et~al.}, ``{Measurement of the
  azimuthal anisotropy of charged-particle production in Xe$+$Xe collisions at
  $\sqrt{s_{_{\rm NN}}} = 5.44$ TeV with the ATLAS detector}'',
  \href{https://doi.org/10.1103/PhysRevC.101.024906}{{\em Phys. Rev. C}
  {\bfseries 101} (2020) 024906},
  \href{https://arxiv.org/abs/1911.04812}{{\ttfamily arXiv:1911.04812
  [nucl-ex]}}.

\bibitem{CMS:2013wjq}
{\bfseries CMS} Collaboration, S.~Chatrchyan {\em et~al.}, ``{Measurement of
  Higher-Order Harmonic Azimuthal Anisotropy in PbPb Collisions at
  $\sqrt{s_{_{\rm NN}}}$ = 2.76 TeV}'',
  \href{https://doi.org/10.1103/PhysRevC.89.044906}{{\em Phys. Rev. C}
  {\bfseries 89} (2014) 044906},
  \href{https://arxiv.org/abs/1310.8651}{{\ttfamily arXiv:1310.8651
  [nucl-ex]}}.

\bibitem{ALICE:2018rtz}
{\bfseries ALICE} Collaboration, S.~Acharya {\em et~al.}, ``{Energy dependence
  and fluctuations of anisotropic flow in Pb--Pb collisions at $
  \sqrt{s_{\mathrm{NN}}}=5.02 $ and 2.76 TeV}'',
  \href{https://doi.org/10.1007/JHEP07(2018)103}{{\em JHEP} {\bfseries 07}
  (2018) 103}, \href{https://arxiv.org/abs/1804.02944}{{\ttfamily
  arXiv:1804.02944 [nucl-ex]}}.

\bibitem{ALICE:2022dtx}
{\bfseries ALICE} Collaboration, S.~Acharya {\em et~al.}, ``{Observation of
  flow angle and flow magnitude fluctuations in Pb-Pb collisions at
  $\sqrt{s_{_{\rm NN}}} =$5.02TeV at the CERN Large Hadron Collider}'',
  \href{https://doi.org/10.1103/PhysRevC.107.L051901}{{\em Phys. Rev. C}
  {\bfseries 107} (2023) L051901},
  \href{https://arxiv.org/abs/2206.04574}{{\ttfamily arXiv:2206.04574
  [nucl-ex]}}.

\bibitem{ALICE:2024fcv}
{\bfseries ALICE} Collaboration, S.~Acharya {\em et~al.}, ``{Systematic study
  of flow vector fluctuations in $\sqrt{s_{_{\rm NN}}} =$5.02 TeV Pb-Pb
  collisions}'', \href{https://doi.org/10.1103/PhysRevC.109.065202}{{\em Phys.
  Rev. C} {\bfseries 109} (2024) 065202},
  \href{https://arxiv.org/abs/2403.15213}{{\ttfamily arXiv:2403.15213
  [nucl-ex]}}.

\bibitem{ALICE:2023tvh}
{\bfseries ALICE} Collaboration, S.~Acharya {\em et~al.}, ``{Pseudorapidity
  dependence of anisotropic flow and its decorrelations using long-range
  multiparticle correlations in Pb\textendash{}Pb and Xe\textendash{}Xe
  collisions}'', \href{https://doi.org/10.1016/j.physletb.2024.138477}{{\em
  Phys. Lett. B} {\bfseries 850} (2024) 138477},
  \href{https://arxiv.org/abs/2307.11116}{{\ttfamily arXiv:2307.11116
  [nucl-ex]}}. [Erratum: Phys.Lett.B 853, 138659 (2024)].

\bibitem{ATLAS:2013xzf}
{\bfseries ATLAS} Collaboration, G.~Aad {\em et~al.}, ``{Measurement of the
  distributions of event-by-event flow harmonics in lead-lead collisions at
  $\sqrt{s_{_{\rm NN}}}$ = 2.76 TeV with the ATLAS detector at the LHC}'',
  \href{https://doi.org/10.1007/JHEP11(2013)183}{{\em JHEP} {\bfseries 11}
  (2013) 183}, \href{https://arxiv.org/abs/1305.2942}{{\ttfamily
  arXiv:1305.2942 [hep-ex]}}.

\bibitem{CMS:2017glf}
{\bfseries CMS} Collaboration, A.~M. Sirunyan {\em et~al.}, ``{Non-Gaussian
  elliptic-flow fluctuations in PbPb collisions at
  $\sqrt{\smash[b]{s_{_\text{NN}}}} = 5.02$ TeV}'',
  \href{https://doi.org/10.1016/j.physletb.2018.11.063}{{\em Phys. Lett. B}
  {\bfseries 789} (2019) 643--665},
  \href{https://arxiv.org/abs/1711.05594}{{\ttfamily arXiv:1711.05594
  [nucl-ex]}}.

\bibitem{ALICE:2016kpq}
{\bfseries ALICE} Collaboration, J.~Adam {\em et~al.}, ``{Correlated
  event-by-event fluctuations of flow harmonics in Pb--Pb collisions at
  $\sqrt{s_{_{\rm NN}}}=2.76$ TeV}'',
  \href{https://doi.org/10.1103/PhysRevLett.117.182301}{{\em Phys. Rev. Lett.}
  {\bfseries 117} (2016) 182301},
  \href{https://arxiv.org/abs/1604.07663}{{\ttfamily arXiv:1604.07663
  [nucl-ex]}}.

\bibitem{ALICE:2017fcd}
{\bfseries ALICE} Collaboration, S.~Acharya {\em et~al.}, ``{Linear and
  non-linear flow modes in Pb--Pb collisions at $\sqrt{s_{_{\rm NN}}} =$ 2.76
  TeV}'', \href{https://doi.org/10.1016/j.physletb.2017.07.060}{{\em Phys.
  Lett. B} {\bfseries 773} (2017) 68--80},
  \href{https://arxiv.org/abs/1705.04377}{{\ttfamily arXiv:1705.04377
  [nucl-ex]}}.

\bibitem{ALICE:2017kwu}
{\bfseries ALICE} Collaboration, S.~Acharya {\em et~al.}, ``{Systematic studies
  of correlations between different order flow harmonics in Pb--Pb collisions
  at $\sqrt{s_{_{\rm NN}}}$ = 2.76 TeV}'',
  \href{https://doi.org/10.1103/PhysRevC.97.024906}{{\em Phys. Rev. C}
  {\bfseries 97} (2018) 024906},
  \href{https://arxiv.org/abs/1709.01127}{{\ttfamily arXiv:1709.01127
  [nucl-ex]}}.

\bibitem{ALICE:2020sup}
{\bfseries ALICE} Collaboration, S.~Acharya {\em et~al.}, ``{Higher harmonic
  non-linear flow modes of charged hadrons in Pb--Pb collisions at
  $\sqrt{s_{\rm{NN}}}$ = 5.02 TeV}'',
  \href{https://doi.org/10.1007/JHEP05(2020)085}{{\em JHEP} {\bfseries 05}
  (2020) 085}, \href{https://arxiv.org/abs/2002.00633}{{\ttfamily
  arXiv:2002.00633 [nucl-ex]}}.

\bibitem{ALICE:2021adw}
{\bfseries ALICE} Collaboration, S.~Acharya {\em et~al.}, ``{Measurements of
  mixed harmonic cumulants in Pb\textendash{}Pb collisions at $\sqrt{s_{_{\rm
  NN}}} = 5.02$ TeV}'',
  \href{https://doi.org/10.1016/j.physletb.2021.136354}{{\em Phys. Lett. B}
  {\bfseries 818} (2021) 136354},
  \href{https://arxiv.org/abs/2102.12180}{{\ttfamily arXiv:2102.12180
  [nucl-ex]}}.

\bibitem{ATLAS:2015qwl}
{\bfseries ATLAS} Collaboration, G.~Aad {\em et~al.}, ``{Measurement of the
  correlation between flow harmonics of different order in lead-lead collisions
  at $\sqrt{s_{_{\rm NN}}} = 2.76$ TeV with the ATLAS detector}'',
  \href{https://doi.org/10.1103/PhysRevC.92.034903}{{\em Phys. Rev. C}
  {\bfseries 92} (2015) 034903},
  \href{https://arxiv.org/abs/1504.01289}{{\ttfamily arXiv:1504.01289
  [hep-ex]}}.

\bibitem{Parkkila:2021tqq}
J.~E. Parkkila, A.~Onnerstad, and D.~J. Kim, ``{Bayesian estimation of the
  specific shear and bulk viscosity of the quark-gluon plasma with additional
  flow harmonic observables}'',
  \href{https://doi.org/10.1103/PhysRevC.104.054904}{{\em Phys. Rev. C}
  {\bfseries 104} (2021) 054904},
  \href{https://arxiv.org/abs/2106.05019}{{\ttfamily arXiv:2106.05019
  [hep-ph]}}.

\bibitem{Niemi:2012aj}
H.~Niemi, G.~S. Denicol, H.~Holopainen, and P.~Huovinen, ``{Event-by-event
  distributions of azimuthal asymmetries in ultrarelativistic heavy-ion
  collisions}'', \href{https://doi.org/10.1103/PhysRevC.87.054901}{{\em Phys.
  Rev. C} {\bfseries 87} (2013) 054901},
  \href{https://arxiv.org/abs/1212.1008}{{\ttfamily arXiv:1212.1008
  [nucl-th]}}.

\bibitem{Song:2010mg}
H.~Song, S.~A. Bass, U.~Heinz, T.~Hirano, and C.~Shen, ``{200 A GeV Au+Au
  collisions serve a nearly perfect quark-gluon liquid}'',
  \href{https://doi.org/10.1103/PhysRevLett.106.192301}{{\em Phys. Rev. Lett.}
  {\bfseries 106} (2011) 192301},
  \href{https://arxiv.org/abs/1011.2783}{{\ttfamily arXiv:1011.2783
  [nucl-th]}}. [Erratum: Phys.Rev.Lett. 109, 139904 (2012)].

\bibitem{Bilandzic:2013kga}
A.~Bilandzic, C.~H. Christensen, K.~Gulbrandsen, A.~Hansen, and Y.~Zhou,
  ``{Generic framework for anisotropic flow analyses with multiparticle
  azimuthal correlations}'',
  \href{https://doi.org/10.1103/PhysRevC.89.064904}{{\em Phys. Rev. C}
  {\bfseries 89} (2014) 064904},
  \href{https://arxiv.org/abs/1312.3572}{{\ttfamily arXiv:1312.3572
  [nucl-ex]}}.

\bibitem{Zhu:2016puf}
X.~Zhu, Y.~Zhou, H.~Xu, and H.~Song, ``{Correlations of flow harmonics in 2.76A
  TeV Pb--Pb collisions}'',
  \href{https://doi.org/10.1103/PhysRevC.95.044902}{{\em Phys. Rev. C}
  {\bfseries 95} (2017) 044902},
  \href{https://arxiv.org/abs/1608.05305}{{\ttfamily arXiv:1608.05305
  [nucl-th]}}.

\bibitem{Hagino:2006fj}
K.~Hagino, N.~W. Lwin, and M.~Yamagami, ``{Deformation parameter for diffuse
  density}'', \href{https://doi.org/10.1103/PhysRevC.74.017310}{{\em Phys. Rev.
  C} {\bfseries 74} (2006) 017310},
  \href{https://arxiv.org/abs/nucl-th/0604048}{{\ttfamily
  arXiv:nucl-th/0604048}}.

\bibitem{Raman:2001nnq}
S.~Raman, C.~W.~G. Nestor, Jr, and P.~Tikkanen, ``{Transition probability from
  the ground to the first-excited $2^{+}$ state of even-even nuclides}'',
  \href{https://doi.org/10.1006/adnd.2001.0858}{{\em Atom. Data Nucl. Data
  Tabl.} {\bfseries 78} (2001) 1--128}.

\bibitem{Pritychenko:2013gwa}
B.~Pritychenko, M.~Birch, B.~Singh, and M.~Horoi, ``{Tables of E2 Transition
  Probabilities from the first $2^{+}$ States in Even-Even Nuclei}'',
  \href{https://doi.org/10.1016/j.adt.2015.10.001}{{\em Atom. Data Nucl. Data
  Tabl.} {\bfseries 107} (2016) 1--139},
  \href{https://arxiv.org/abs/1312.5975}{{\ttfamily arXiv:1312.5975
  [nucl-th]}}. [Erratum: Atom.Data Nucl.Data Tabl. 114, 371--374 (2017)].

\bibitem{Henderson:2020yql}
J.~Henderson, ``{Convergence of electric quadrupole rotational invariants from
  the nuclear shell model}'',
  \href{https://doi.org/10.1103/PhysRevC.102.054306}{{\em Phys. Rev. C}
  {\bfseries 102} (2020) 054306},
  \href{https://arxiv.org/abs/2005.11210}{{\ttfamily arXiv:2005.11210
  [nucl-th]}}.

\bibitem{Jia:2022ozr}
J.~Jia {\em et~al.}, ``{Imaging the initial condition of heavy-ion collisions
  and nuclear structure across the nuclide chart}'',
  \href{https://doi.org/10.1007/s41365-024-01589-w}{{\em Nucl. Sci. Tech.}
  {\bfseries 35} (2024) 220},
  \href{https://arxiv.org/abs/2209.11042}{{\ttfamily arXiv:2209.11042
  [nucl-ex]}}.

\bibitem{Bilandzic:2010jr}
A.~Bilandzic, R.~Snellings, and S.~Voloshin, ``{Flow analysis with cumulants:
  Direct calculations}'',
  \href{https://doi.org/10.1103/PhysRevC.83.044913}{{\em Phys. Rev. C}
  {\bfseries 83} (2011) 044913},
  \href{https://arxiv.org/abs/1010.0233}{{\ttfamily arXiv:1010.0233
  [nucl-ex]}}.

\bibitem{Borghini:2000sa}
N.~Borghini, P.~M. Dinh, and J.-Y. Ollitrault, ``{A New method for measuring
  azimuthal distributions in nucleus-nucleus collisions}'',
  \href{https://doi.org/10.1103/PhysRevC.63.054906}{{\em Phys. Rev. C}
  {\bfseries 63} (2001) 054906},
  \href{https://arxiv.org/abs/nucl-th/0007063}{{\ttfamily
  arXiv:nucl-th/0007063}}.

\bibitem{Moravcova:2020wnf}
Z.~Moravcova, K.~Gulbrandsen, and Y.~Zhou, ``{Generic algorithm for
  multiparticle cumulants of azimuthal correlations in high energy nucleus
  collisions}'', \href{https://doi.org/10.1103/PhysRevC.103.024913}{{\em Phys.
  Rev. C} {\bfseries 103} (2021) 024913},
  \href{https://arxiv.org/abs/2005.07974}{{\ttfamily arXiv:2005.07974
  [nucl-th]}}.

\bibitem{Voloshin:2007pc}
S.~A. Voloshin, A.~M. Poskanzer, A.~Tang, and G.~Wang, ``{Elliptic flow in the
  Gaussian model of eccentricity fluctuations}'',
  \href{https://doi.org/10.1016/j.physletb.2007.11.043}{{\em Phys. Lett. B}
  {\bfseries 659} (2008) 537--541},
  \href{https://arxiv.org/abs/0708.0800}{{\ttfamily arXiv:0708.0800
  [nucl-th]}}.

\bibitem{Alver:2010gr}
B.~Alver and G.~Roland, ``{Collision geometry fluctuations and triangular flow
  in heavy-ion collisions}'',
  \href{https://doi.org/10.1103/PhysRevC.82.039903}{{\em Phys. Rev. C}
  {\bfseries 81} (2010) 054905},
  \href{https://arxiv.org/abs/1003.0194}{{\ttfamily arXiv:1003.0194
  [nucl-th]}}. [Erratum: Phys.Rev.C 82, 039903 (2010)].

\bibitem{Bhalerao:2014xra}
R.~S. Bhalerao, J.-Y. Ollitrault, and S.~Pal, ``{Characterizing flow
  fluctuations with moments}'',
  \href{https://doi.org/10.1016/j.physletb.2015.01.019}{{\em Phys. Lett. B}
  {\bfseries 742} (2015) 94--98},
  \href{https://arxiv.org/abs/1411.5160}{{\ttfamily arXiv:1411.5160
  [nucl-th]}}.

\bibitem{Bhalerao:2013ina}
R.~S. Bhalerao, J.-Y. Ollitrault, and S.~Pal, ``{Event-plane correlators}'',
  \href{https://doi.org/10.1103/PhysRevC.88.024909}{{\em Phys. Rev. C}
  {\bfseries 88} (2013) 024909},
  \href{https://arxiv.org/abs/1307.0980}{{\ttfamily arXiv:1307.0980
  [nucl-th]}}.

\bibitem{Yan:2015jma}
L.~Yan and J.-Y. Ollitrault, ``{$v_4, v_5, v_6, v_7$: nonlinear hydrodynamic
  response versus LHC data}'',
  \href{https://doi.org/10.1016/j.physletb.2015.03.040}{{\em Phys. Lett. B}
  {\bfseries 744} (2015) 82--87},
  \href{https://arxiv.org/abs/1502.02502}{{\ttfamily arXiv:1502.02502
  [nucl-th]}}.

\bibitem{Zhou:2015eya}
Y.~Zhou, K.~Xiao, Z.~Feng, F.~Liu, and R.~Snellings, ``{Anisotropic
  distributions in a multiphase transport model}'',
  \href{https://doi.org/10.1103/PhysRevC.93.034909}{{\em Phys. Rev. C}
  {\bfseries 93} (2016) 034909},
  \href{https://arxiv.org/abs/1508.03306}{{\ttfamily arXiv:1508.03306
  [nucl-ex]}}.

\bibitem{Qian:2016fpi}
J.~Qian, U.~W. Heinz, and J.~Liu, ``{Mode-coupling effects in anisotropic flow
  in heavy-ion collisions}'',
  \href{https://doi.org/10.1103/PhysRevC.93.064901}{{\em Phys. Rev. C}
  {\bfseries 93} (2016) 064901},
  \href{https://arxiv.org/abs/1602.02813}{{\ttfamily arXiv:1602.02813
  [nucl-th]}}.

\bibitem{Parkkila:2021yha}
J.~E. Parkkila, A.~Onnerstad, S.~F. Taghavi, C.~Mordasini, A.~Bilandzic,
  M.~Virta, and D.~J. Kim, ``{New constraints for QCD matter from improved
  Bayesian parameter estimation in heavy-ion collisions at LHC}'',
  \href{https://doi.org/10.1016/j.physletb.2022.137485}{{\em Phys. Lett. B}
  {\bfseries 835} (2022) 137485},
  \href{https://arxiv.org/abs/2111.08145}{{\ttfamily arXiv:2111.08145
  [hep-ph]}}.

\bibitem{Huo:2017nms}
P.~Huo, K.~Gajdo\v{s}ov\'a, J.~Jia, and Y.~Zhou, ``{Importance of non-flow in
  mixed-harmonic multi-particle correlations in small collision systems}'',
  \href{https://doi.org/10.1016/j.physletb.2017.12.035}{{\em Phys. Lett. B}
  {\bfseries 777} (2018) 201--206},
  \href{https://arxiv.org/abs/1710.07567}{{\ttfamily arXiv:1710.07567
  [nucl-ex]}}.

\bibitem{ALICE:2014dwt}
{\bfseries ALICE} Collaboration, B.~B. Abelev {\em et~al.}, ``{Multiparticle
  azimuthal correlations in p--Pb and Pb--Pb collisions at the CERN Large
  Hadron Collider}'', \href{https://doi.org/10.1103/PhysRevC.90.054901}{{\em
  Phys. Rev. C} {\bfseries 90} (2014) 054901},
  \href{https://arxiv.org/abs/1406.2474}{{\ttfamily arXiv:1406.2474
  [nucl-ex]}}.

\bibitem{ALICE:2008ngc}
{\bfseries ALICE} Collaboration, K.~Aamodt {\em et~al.}, ``{The ALICE
  experiment at the CERN LHC}'',
  \href{https://doi.org/10.1088/1748-0221/3/08/S08002}{{\em JINST} {\bfseries
  3} (2008) S08002}.

\bibitem{ALICE:2004fvi}
{\bfseries ALICE} Collaboration, P.~Cortese {\em et~al.}, ``{ALICE: Physics
  performance report, volume I}'',
  \href{https://doi.org/10.1088/0954-3899/30/11/001}{{\em J. Phys. G}
  {\bfseries 30} (2004) 1517--1763}.

\bibitem{ALICE:2005vhb}
{\bfseries ALICE} Collaboration, C.~W. Fabjan {\em et~al.}, ``{ALICE: Physics
  Performance Report, volume II}'',
  \href{https://doi.org/10.1088/0954-3899/32/10/001}{{\em J. Phys. G}
  {\bfseries 32} (2006) 1295--2040}.

\bibitem{ALICE:2014sbx}
{\bfseries ALICE} Collaboration, B.~B. Abelev {\em et~al.}, ``{Performance of
  the ALICE Experiment at the CERN LHC}'',
  \href{https://doi.org/10.1142/S0217751X14300440}{{\em Int. J. Mod. Phys. A}
  {\bfseries 29} (2014) 1430044},
  \href{https://arxiv.org/abs/1402.4476}{{\ttfamily arXiv:1402.4476
  [nucl-ex]}}.

\bibitem{ALICE:2013axi}
{\bfseries ALICE} Collaboration, E.~Abbas {\em et~al.}, ``{Performance of the
  ALICE VZERO system}'',
  \href{https://doi.org/10.1088/1748-0221/8/10/P10016}{{\em JINST} {\bfseries
  8} (2013) P10016}, \href{https://arxiv.org/abs/1306.3130}{{\ttfamily
  arXiv:1306.3130 [nucl-ex]}}.

\bibitem{ALICE:2010tia}
{\bfseries ALICE} Collaboration, K.~Aamodt {\em et~al.}, ``{Alignment of the
  ALICE Inner Tracking System with cosmic-ray tracks}'',
  \href{https://doi.org/10.1088/1748-0221/5/03/P03003}{{\em JINST} {\bfseries
  5} (2010) P03003}, \href{https://arxiv.org/abs/1001.0502}{{\ttfamily
  arXiv:1001.0502 [physics.ins-det]}}.

\bibitem{Alme:2010ke}
J.~Alme {\em et~al.}, ``{The ALICE TPC, a large 3-dimensional tracking device
  with fast readout for ultra-high multiplicity events}'',
  \href{https://doi.org/10.1016/j.nima.2010.04.042}{{\em Nucl. Instrum. Meth.
  A} {\bfseries 622} (2010) 316--367},
  \href{https://arxiv.org/abs/1001.1950}{{\ttfamily arXiv:1001.1950
  [physics.ins-det]}}.

\bibitem{ALICE:2013hur}
{\bfseries ALICE} Collaboration, B.~Abelev {\em et~al.}, ``{Centrality
  determination of Pb--Pb collisions at $\sqrt{s_{_{\rm NN}}}$ = 2.76 TeV with
  ALICE}'', \href{https://doi.org/10.1103/PhysRevC.88.044909}{{\em Phys. Rev.
  C} {\bfseries 88} (2013) 044909},
  \href{https://arxiv.org/abs/1301.4361}{{\ttfamily arXiv:1301.4361
  [nucl-ex]}}.

\bibitem{Wang:1991hta}
X.-N. Wang and M.~Gyulassy, ``{HIJING: A Monte Carlo model for multiple jet
  production in p p, p A and A A collisions}'',
  \href{https://doi.org/10.1103/PhysRevD.44.3501}{{\em Phys. Rev. D} {\bfseries
  44} (1991) 3501--3516}.

\bibitem{Gyulassy:1994ew}
M.~Gyulassy and X.-N. Wang, ``{HIJING 1.0: A Monte Carlo program for parton and
  particle production in high-energy hadronic and nuclear collisions}'',
  \href{https://doi.org/10.1016/0010-4655(94)90057-4}{{\em Comput. Phys.
  Commun.} {\bfseries 83} (1994) 307},
  \href{https://arxiv.org/abs/nucl-th/9502021}{{\ttfamily
  arXiv:nucl-th/9502021}}.

\bibitem{Brun:1994aa}
R.~Brun {\em et~al.}, \href{https://doi.org/10.17181/CERN.MUHF.DMJ1}{{\em
  {GEANT: Detector Description and Simulation Tool}}}.
\newblock CERN Program Library. CERN, Geneva, 1993.
\newblock \url{https://cds.cern.ch/record/1082634}.
\newblock Long Writeup W5013.

\bibitem{Barlow:2002yb}
R.~Barlow, ``{Systematic errors: Facts and fictions}'', in {\em {Conference on
  Advanced Statistical Techniques in Particle Physics}}, pp.~134--144.
\newblock 7, 2002.
\newblock \href{https://arxiv.org/abs/hep-ex/0207026}{{\ttfamily
  arXiv:hep-ex/0207026}}.

\bibitem{Schenke:2020mbo}
B.~Schenke, C.~Shen, and P.~Tribedy, ``{Running the gamut of high energy
  nuclear collisions}'',
  \href{https://doi.org/10.1103/PhysRevC.102.044905}{{\em Phys. Rev. C}
  {\bfseries 102} (2020) 044905},
  \href{https://arxiv.org/abs/2005.14682}{{\ttfamily arXiv:2005.14682
  [nucl-th]}}.

\bibitem{Mantysaari:2022ffw}
H.~M\"antysaari, B.~Schenke, C.~Shen, and W.~Zhao, ``{Bayesian inference of the
  fluctuating proton shape}'',
  \href{https://doi.org/10.1016/j.physletb.2022.137348}{{\em Phys. Lett. B}
  {\bfseries 833} (2022) 137348},
  \href{https://arxiv.org/abs/2202.01998}{{\ttfamily arXiv:2202.01998
  [hep-ph]}}.

\bibitem{Liu:2023pav}
Q.~Liu, S.~Zhao, H.-j. Xu, and H.~Song, ``{Determining the neutron skin
  thickness by relativistic semi-isobaric collisions}'',
  \href{https://doi.org/10.1103/PhysRevC.109.034912}{{\em Phys. Rev. C}
  {\bfseries 109} (2024) 034912},
  \href{https://arxiv.org/abs/2311.01747}{{\ttfamily arXiv:2311.01747
  [nucl-th]}}.

\bibitem{STAR:2022gki}
{\bfseries STAR} Collaboration, M.~Abdallah {\em et~al.}, ``{Collision-System
  and Beam-Energy Dependence of Anisotropic Flow Fluctuations}'',
  \href{https://doi.org/10.1103/PhysRevLett.129.252301}{{\em Phys. Rev. Lett.}
  {\bfseries 129} (2022) 252301},
  \href{https://arxiv.org/abs/2201.10365}{{\ttfamily arXiv:2201.10365
  [nucl-ex]}}.

\bibitem{Molnar:2008xj}
D.~Molnar and P.~Huovinen, ``{Dissipative effects from transport and viscous
  hydrodynamics}'', \href{https://doi.org/10.1088/0954-3899/35/10/104125}{{\em
  J. Phys. G} {\bfseries 35} (2008) 104125},
  \href{https://arxiv.org/abs/0806.1367}{{\ttfamily arXiv:0806.1367
  [nucl-th]}}.

\bibitem{Song:2008si}
H.~Song and U.~W. Heinz, ``{Multiplicity scaling in ideal and viscous
  hydrodynamics}'', \href{https://doi.org/10.1103/PhysRevC.78.024902}{{\em
  Phys. Rev. C} {\bfseries 78} (2008) 024902},
  \href{https://arxiv.org/abs/0805.1756}{{\ttfamily arXiv:0805.1756
  [nucl-th]}}.

\bibitem{Dimri:2023wup}
A.~Dimri, S.~Bhatta, and J.~Jia, ``{Impact of nuclear shape fluctuations in
  high-energy heavy ion collisions}'',
  \href{https://doi.org/10.1140/epja/s10050-023-00965-1}{{\em Eur. Phys. J. A}
  {\bfseries 59} (2023) 45}, \href{https://arxiv.org/abs/2301.03556}{{\ttfamily
  arXiv:2301.03556 [nucl-th]}}.

\bibitem{Moller:2015fba}
P.~M\"oller, A.~J. Sierk, T.~Ichikawa, and H.~Sagawa, ``{Nuclear ground-state
  masses and deformations: FRDM(2012)}'',
  \href{https://doi.org/10.1016/j.adt.2015.10.002}{{\em Atom. Data Nucl. Data
  Tabl.} {\bfseries 109-110} (2016) 1--204},
  \href{https://arxiv.org/abs/1508.06294}{{\ttfamily arXiv:1508.06294
  [nucl-th]}}.

\bibitem{Giacalone:2017dud}
G.~Giacalone, J.~Noronha-Hostler, M.~Luzum, and J.-Y. Ollitrault,
  ``{Hydrodynamic predictions for 5.44 TeV Xe+Xe collisions}'',
  \href{https://doi.org/10.1103/PhysRevC.97.034904}{{\em Phys. Rev. C}
  {\bfseries 97} (2018) 034904},
  \href{https://arxiv.org/abs/1711.08499}{{\ttfamily arXiv:1711.08499
  [nucl-th]}}.

\bibitem{Alver:2010dn}
B.~H. Alver, C.~Gombeaud, M.~Luzum, and J.-Y. Ollitrault, ``{Triangular flow in
  hydrodynamics and transport theory}'',
  \href{https://doi.org/10.1103/PhysRevC.82.034913}{{\em Phys. Rev. C}
  {\bfseries 82} (2010) 034913},
  \href{https://arxiv.org/abs/1007.5469}{{\ttfamily arXiv:1007.5469
  [nucl-th]}}.

\bibitem{ALICE:2021wim}
{\bfseries ALICE} Collaboration, S.~Acharya {\em et~al.}, ``{ALICE physics
  projections for a short oxygen-beam run at the LHC}'',
  \href{https://arxiv.org/abs/ALICE-PUBLIC-2021-004}{{\ttfamily
  ALICE-PUBLIC-2021-004}}. \url{https://cds.cern.ch/record/2765973}.

\bibitem{Zhang:2024vkh}
C.~Zhang, J.~Chen, G.~Giacalone, S.~Huang, J.~Jia, and Y.-G. Ma,
  ``{$Ab$-$initio$ nucleon-nucleon correlations and their impact on high energy
  $^{16}$O+$^{16}$O collisions}'',
  \href{https://arxiv.org/abs/2404.08385}{{\ttfamily arXiv:2404.08385
  [nucl-th]}}.

\bibitem{YuanyuanWang:2024sgp}
Y.~Wang, S.~Zhao, B.~Cao, H.-j. Xu, and H.~Song, ``{Exploring the compactness
  of \ensuremath{\alpha} clusters in O16 nuclei with relativistic O16+O16
  collisions}'', \href{https://doi.org/10.1103/PhysRevC.109.L051904}{{\em Phys.
  Rev. C} {\bfseries 109} (2024) L051904},
  \href{https://arxiv.org/abs/2401.15723}{{\ttfamily arXiv:2401.15723
  [nucl-th]}}.

\bibitem{Bally:2022vgo}
B.~Bally {\em et~al.}, ``{Imaging the initial condition of heavy-ion collisions
  and nuclear structure across the nuclide chart}'',
  \href{https://arxiv.org/abs/2209.11042}{{\ttfamily arXiv:2209.11042
  [nucl-ex]}}.

\end{thebibliography}\endgroup

\newpage
\appendix

\section{Supplemental Material}
\label{app:SM}
This section present measurements of $v_4^{\rm L}$, $\chi_{4,22}$, NSC$(3,2)$ that are less sensitive to the deformation and diffuseness parameters.

\begin{figure}[!htb]
    \begin{center}
      \includegraphics[width=0.8\textwidth]{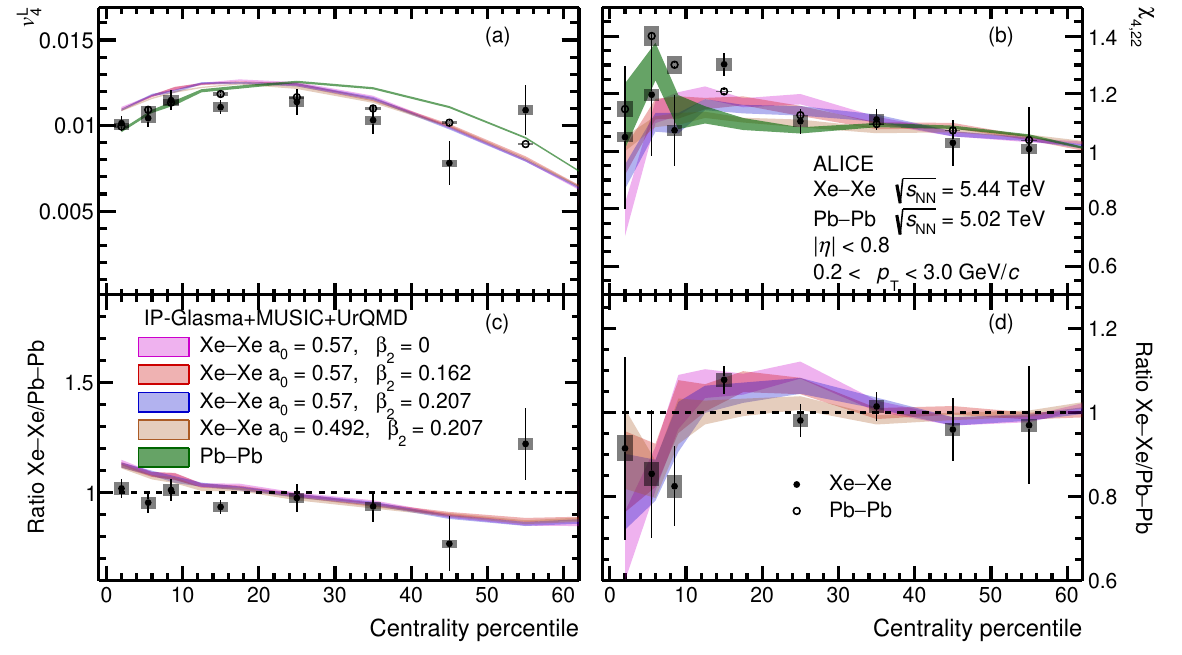}
      \caption[.]{Panels (a) and (b): Charged particle $v_{4}^{\rm L}$ (left) and $\chi_{4,22}$ (right) as a function of centrality in \XeXe and \PbPb collisions at \fivefourfour and \fivenn, respectively. Panels (c) and (d): Ratio between \XeXe and \PbPb $v_{4}^{\rm L}$ (left) and $\chi_{4,22}$ (right). Statistical and systematical uncertainties are shown as vertical lines and grey boxes, respectively. The measurements are compared with IP-Glasma+MUSIC+UrQMD calculations~\cite{Schenke:2020mbo,Mantysaari:2022ffw}. The thickness of the bands represent statistical uncertainties.}
      \label{fig:NonSensitive}
    \end{center}
\end{figure}
Figure~\ref{fig:NonSensitive} shows the centrality dependence of $v_{4}^{\rm L}$and $\chi_{4,22}$ in \XeXe and \PbPb collisions.
In the upper panels, $v_{4}^{\rm L}$ increases with centrality percentile up to 20\% and thereafter decreases towards more peripheral collisions, while $\chi_{4,22}$ measurements in both \XeXe and \PbPb collisions exhibit a modest centrality dependence, with hints of a finer structure in the most central collisions. 
The IP-Glasma+MUSIC+UrQMD calculations for $v_{3}\{2, |\Delta\eta| > 0.8\}$ show no variation with $\beta_2$. A similar lack of dependence is also observed for $\chi_{4,22}$.

\begin{figure}[!htb]
    \begin{center}
      \includegraphics[width=0.4\textwidth]{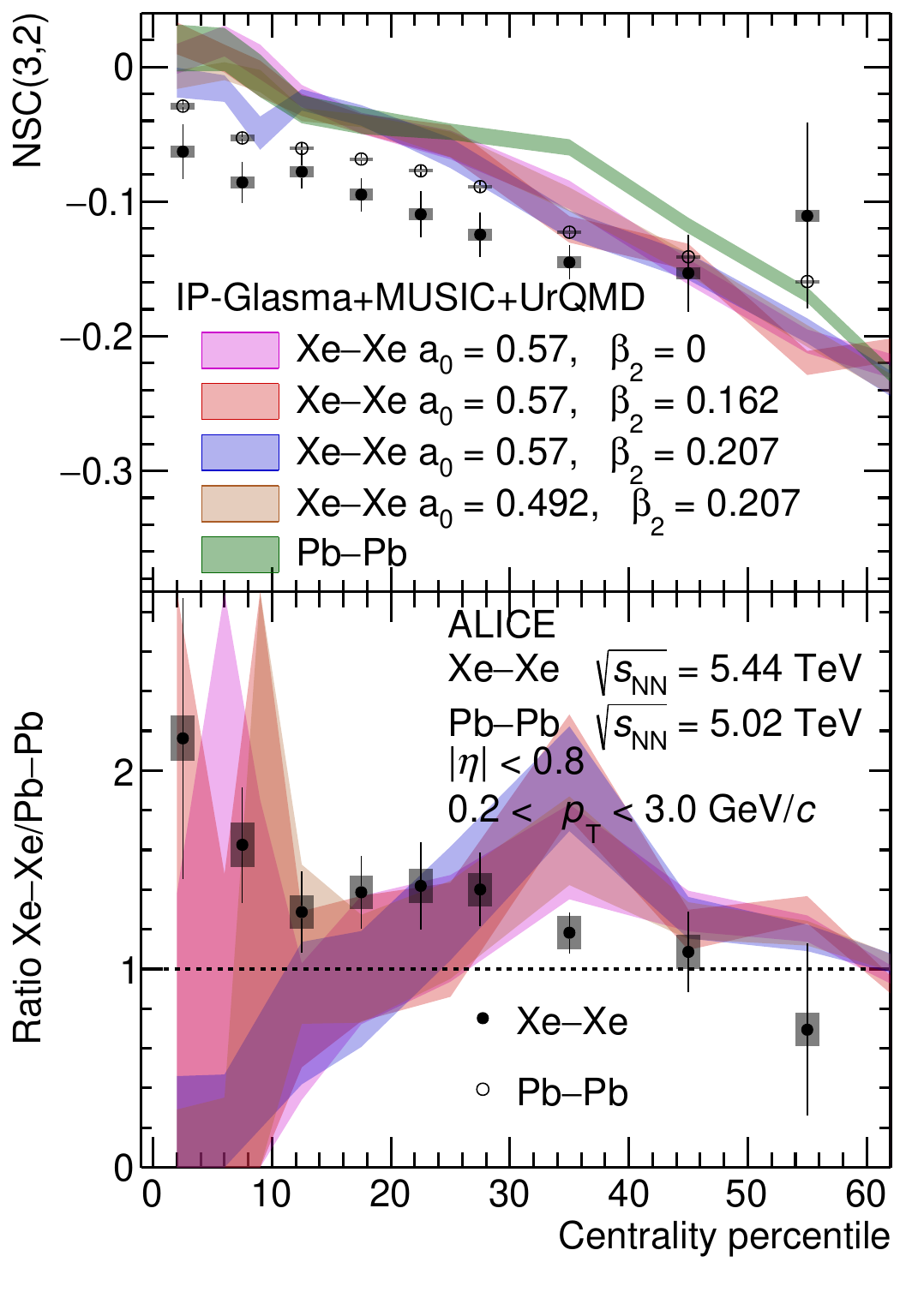}
      \caption[.]{Top Panel: Charged particle 
NSC$(3,2)$ as a function of centrality in \XeXe and \PbPb collisions at \fivefourfour and \fivenn, respectively. Bottom Panel: Ratio between \XeXe and \PbPb NSC$(3,2)$. Statistical and systematical uncertainties are shown as vertical lines and grey boxes, respectively. The measurements are compared with IP-Glasma+MUSIC+UrQMD calculations~\cite{Schenke:2020mbo,Mantysaari:2022ffw} to constrain the $\beta_2$ and $a_0$ parameters of $^{129}$Xe nuclei. The thickness of the bands represent statistical uncertainties.}
      \label{fig:NSC32}
    \end{center}
\end{figure}

Figure~\ref{fig:NSC32} shows the centrality dependence of NSC$(3,2)$~\cite{Bilandzic:2013kga} in \XeXe and \PbPb collisions. The upper panel of Fig.~\ref{fig:NSC32} shows that NSC$(3,2)$ decreases from central to peripheral collisions in both collision systems, with the results in \PbPb generally larger than those in \XeXe. The lower panel presents the NSC$(3,2)$ ratio between \XeXe and \PbPb, which decreases from central to peripheral collisions for the presented centrality range. The sizeable statistical uncertainties in both measurements and IP-Glasma+MUSIC+UrQMD calculations prevent drawing a firm conclusion on which set of nuclear structure parameters is preferred. 

%
%

\clearpage
\section{The ALICE Collaboration}
\label{app:collab}
\begin{flushleft} 
\small

S.~Acharya\,\orcidlink{0000-0002-9213-5329}\,$^{\rm 127}$, 
A.~Agarwal$^{\rm 135}$, 
G.~Aglieri Rinella\,\orcidlink{0000-0002-9611-3696}\,$^{\rm 32}$, 
L.~Aglietta\,\orcidlink{0009-0003-0763-6802}\,$^{\rm 24}$, 
M.~Agnello\,\orcidlink{0000-0002-0760-5075}\,$^{\rm 29}$, 
N.~Agrawal\,\orcidlink{0000-0003-0348-9836}\,$^{\rm 25}$, 
Z.~Ahammed\,\orcidlink{0000-0001-5241-7412}\,$^{\rm 135}$, 
S.~Ahmad\,\orcidlink{0000-0003-0497-5705}\,$^{\rm 15}$, 
S.U.~Ahn\,\orcidlink{0000-0001-8847-489X}\,$^{\rm 71}$, 
I.~Ahuja\,\orcidlink{0000-0002-4417-1392}\,$^{\rm 36}$
% ...

\section*{Affiliation Notes}

$^{\rm I}$ Deceased\\
$^{\rm II}$ Also at: Max-Planck-Institut fur Physik, Munich, Germany\\
$^{\rm III}$ Also at: Italian National Agency for New Technologies, Energy and Sustainable Economic Development (ENEA), Bologna, Italy\\
$^{\rm IV}$ Also at: Dipartimento DET del Politecnico di Torino, Turin, Italy\\
$^{\rm V}$ Also at: Department of Applied Physics, Aligarh Muslim University, Aligarh, India\\
$^{\rm VI}$ Also at: Institute of Theoretical Physics, University of Wroclaw, Poland\\
$^{\rm VII}$ Also at: Facultad de Ciencias, Universidad Nacional Aut\'{o}noma de M\'{e}xico, Mexico City, Mexico\\
$^{\rm VIII}$ Also at: An institution covered by a cooperation agreement with CERN\\

\section*{Collaboration Institutes}

$^{1}$ A.I. Alikhanyan National Science Laboratory (Yerevan Physics Institute) Foundation, Yerevan, Armenia\\
$^{2}$ AGH University of Krakow, Cracow, Poland\\
$^{3}$ Bogolyubov Institute for Theoretical Physics, National Academy of Sciences of Ukraine, Kiev, Ukraine\\
$^{4}$ Bose Institute, Department of Physics  and Centre for Astroparticle Physics and Space Science (CAPSS), Kolkata, India\\
$^{5}$ California Polytechnic State University, San Luis Obispo, California, United States\\
$^{6}$ Central China Normal University, Wuhan, China\\
$^{7}$ Centro de Aplicaciones Tecnol\'{o}gicas y Desarrollo Nuclear (CEADEN), Havana, Cuba\\
$^{8}$ Centro de Investigaci\'{o}n y de Estudios Avanzados (CINVESTAV), Mexico City and M\'{e}rida, Mexico\\
$^{9}$ Chicago State University, Chicago, Illinois, United States\\
$^{10}$ China Nuclear Data Center, China Institute of Atomic Energy, Beijing, China\\
$^{11}$ China University of Geosciences, Wuhan, China\\
$^{12}$ Chungbuk National University, Cheongju, Republic of Korea\\
$^{13}$ Comenius University Bratislava, Faculty of Mathematics, Physics and Informatics, Bratislava, Slovak Republic\\
$^{14}$ Creighton University, Omaha, Nebraska, United States\\
$^{15}$ Department of Physics, Aligarh Muslim University, Aligarh, India\\
$^{16}$ Department of Physics, Pusan National University, Pusan, Republic of Korea\\
$^{17}$ Department of Physics, Sejong University, Seoul, Republic of Korea\\
$^{18}$ Department of Physics, University of California, Berkeley, California, United States\\
$^{19}$ Department of Physics, University of Oslo, Oslo, Norway\\
$^{20}$ Department of Physics and Technology, University of Bergen, Bergen, Norway\\
$^{21}$ Dipartimento di Fisica, Universit\`{a} di Pavia, Pavia, Italy\\
$^{22}$ Dipartimento di Fisica dell'Universit\`{a} and Sezione INFN, Cagliari, Italy\\
$^{23}$ Dipartimento di Fisica dell'Universit\`{a} and Sezione INFN, Trieste, Italy\\
$^{24}$ Dipartimento di Fisica dell'Universit\`{a} and Sezione INFN, Turin, Italy\\
$^{25}$ Dipartimento di Fisica e Astronomia dell'Universit\`{a} and Sezione INFN, Bologna, Italy\\
$^{26}$ Dipartimento di Fisica e Astronomia dell'Universit\`{a} and Sezione INFN, Catania, Italy\\
$^{27}$ Dipartimento di Fisica e Astronomia dell'Universit\`{a} and Sezione INFN, Padova, Italy\\
$^{28}$ Dipartimento di Fisica `E.R.~Caianiello' dell'Universit\`{a} and Gruppo Collegato INFN, Salerno, Italy\\
$^{29}$ Dipartimento DISAT del Politecnico and Sezione INFN, Turin, Italy\\
$^{30}$ Dipartimento di Scienze MIFT, Universit\`{a} di Messina, Messina, Italy\\
$^{31}$ Dipartimento Interateneo di Fisica `M.~Merlin' and Sezione INFN, Bari, Italy\\
$^{32}$ European Organization for Nuclear Research (CERN), Geneva, Switzerland\\
$^{33}$ Faculty of Electrical Engineering, Mechanical Engineering and Naval Architecture, University of Split, Split, Croatia\\
$^{34}$ Faculty of Nuclear Sciences and Physical Engineering, Czech Technical University in Prague, Prague, Czech Republic\\
$^{35}$ Faculty of Physics, Sofia University, Sofia, Bulgaria\\
$^{36}$ Faculty of Science, P.J.~\v{S}af\'{a}rik University, Ko\v{s}ice, Slovak Republic\\
$^{37}$ Faculty of Technology, Environmental and Social Sciences, Bergen, Norway\\
$^{38}$ Frankfurt Institute for Advanced Studies, Johann Wolfgang Goethe-Universit\"{a}t Frankfurt, Frankfurt, Germany\\
$^{39}$ Fudan University, Shanghai, China\\
$^{40}$ Gangneung-Wonju National University, Gangneung, Republic of Korea\\
$^{41}$ Gauhati University, Department of Physics, Guwahati, India\\
$^{42}$ Helmholtz-Institut f\"{u}r Strahlen- und Kernphysik, Rheinische Friedrich-Wilhelms-Universit\"{a}t Bonn, Bonn, Germany\\
$^{43}$ Helsinki Institute of Physics (HIP), Helsinki, Finland\\
$^{44}$ High Energy Physics Group,  Universidad Aut\'{o}noma de Puebla, Puebla, Mexico\\
$^{45}$ Horia Hulubei National Institute of Physics and Nuclear Engineering, Bucharest, Romania\\
$^{46}$ HUN-REN Wigner Research Centre for Physics, Budapest, Hungary\\
$^{47}$ Indian Institute of Technology Bombay (IIT), Mumbai, India\\
$^{48}$ Indian Institute of Technology Indore, Indore, India\\
$^{49}$ INFN, Laboratori Nazionali di Frascati, Frascati, Italy\\
$^{50}$ INFN, Sezione di Bari, Bari, Italy\\
$^{51}$ INFN, Sezione di Bologna, Bologna, Italy\\
$^{52}$ INFN, Sezione di Cagliari, Cagliari, Italy\\
$^{53}$ INFN, Sezione di Catania, Catania, Italy\\
$^{54}$ INFN, Sezione di Padova, Padova, Italy\\
$^{55}$ INFN, Sezione di Pavia, Pavia, Italy\\
$^{56}$ INFN, Sezione di Torino, Turin, Italy\\
$^{57}$ INFN, Sezione di Trieste, Trieste, Italy\\
$^{58}$ Inha University, Incheon, Republic of Korea\\
$^{59}$ Institute for Gravitational and Subatomic Physics (GRASP), Utrecht University/Nikhef, Utrecht, Netherlands\\
$^{60}$ Institute of Experimental Physics, Slovak Academy of Sciences, Ko\v{s}ice, Slovak Republic\\
$^{61}$ Institute of Physics, Homi Bhabha National Institute, Bhubaneswar, India\\
$^{62}$ Institute of Physics of the Czech Academy of Sciences, Prague, Czech Republic\\
$^{63}$ Institute of Space Science (ISS), Bucharest, Romania\\
$^{64}$ Institut f\"{u}r Kernphysik, Johann Wolfgang Goethe-Universit\"{a}t Frankfurt, Frankfurt, Germany\\
$^{65}$ Instituto de Ciencias Nucleares, Universidad Nacional Aut\'{o}noma de M\'{e}xico, Mexico City, Mexico\\
$^{66}$ Instituto de F\'{i}sica, Universidade Federal do Rio Grande do Sul (UFRGS), Porto Alegre, Brazil\\
$^{67}$ Instituto de F\'{\i}sica, Universidad Nacional Aut\'{o}noma de M\'{e}xico, Mexico City, Mexico\\
$^{68}$ iThemba LABS, National Research Foundation, Somerset West, South Africa\\
$^{69}$ Jeonbuk National University, Jeonju, Republic of Korea\\
$^{70}$ Johann-Wolfgang-Goethe Universit\"{a}t Frankfurt Institut f\"{u}r Informatik, Fachbereich Informatik und Mathematik, Frankfurt, Germany\\
$^{71}$ Korea Institute of Science and Technology Information, Daejeon, Republic of Korea\\
$^{72}$ KTO Karatay University, Konya, Turkey\\
$^{73}$ Laboratoire de Physique Subatomique et de Cosmologie, Universit\'{e} Grenoble-Alpes, CNRS-IN2P3, Grenoble, France\\
$^{74}$ Lawrence Berkeley National Laboratory, Berkeley, California, United States\\
$^{75}$ Lund University Department of Physics, Division of Particle Physics, Lund, Sweden\\
$^{76}$ Nagasaki Institute of Applied Science, Nagasaki, Japan\\
$^{77}$ Nara Women{'}s University (NWU), Nara, Japan\\
$^{78}$ National and Kapodistrian University of Athens, School of Science, Department of Physics , Athens, Greece\\
$^{79}$ National Centre for Nuclear Research, Warsaw, Poland\\
$^{80}$ National Institute of Science Education and Research, Homi Bhabha National Institute, Jatni, India\\
$^{81}$ National Nuclear Research Center, Baku, Azerbaijan\\
$^{82}$ National Research and Innovation Agency - BRIN, Jakarta, Indonesia\\
$^{83}$ Niels Bohr Institute, University of Copenhagen, Copenhagen, Denmark\\
$^{84}$ Nikhef, National institute for subatomic physics, Amsterdam, Netherlands\\
$^{85}$ Nuclear Physics Group, STFC Daresbury Laboratory, Daresbury, United Kingdom\\
$^{86}$ Nuclear Physics Institute of the Czech Academy of Sciences, Husinec-\v{R}e\v{z}, Czech Republic\\
$^{87}$ Oak Ridge National Laboratory, Oak Ridge, Tennessee, United States\\
$^{88}$ Ohio State University, Columbus, Ohio, United States\\
$^{89}$ Physics department, Faculty of science, University of Zagreb, Zagreb, Croatia\\
$^{90}$ Physics Department, Panjab University, Chandigarh, India\\
$^{91}$ Physics Department, University of Jammu, Jammu, India\\
$^{92}$ Physics Program and International Institute for Sustainability with Knotted Chiral Meta Matter (WPI-SKCM$^{2}$), Hiroshima University, Hiroshima, Japan\\
$^{93}$ Physikalisches Institut, Eberhard-Karls-Universit\"{a}t T\"{u}bingen, T\"{u}bingen, Germany\\
$^{94}$ Physikalisches Institut, Ruprecht-Karls-Universit\"{a}t Heidelberg, Heidelberg, Germany\\
$^{95}$ Physik Department, Technische Universit\"{a}t M\"{u}nchen, Munich, Germany\\
$^{96}$ Politecnico di Bari and Sezione INFN, Bari, Italy\\
$^{97}$ Research Division and ExtreMe Matter Institute EMMI, GSI Helmholtzzentrum f\"ur Schwerionenforschung GmbH, Darmstadt, Germany\\
$^{98}$ Saga University, Saga, Japan\\
$^{99}$ Saha Institute of Nuclear Physics, Homi Bhabha National Institute, Kolkata, India\\
$^{100}$ School of Physics and Astronomy, University of Birmingham, Birmingham, United Kingdom\\
$^{101}$ Secci\'{o}n F\'{\i}sica, Departamento de Ciencias, Pontificia Universidad Cat\'{o}lica del Per\'{u}, Lima, Peru\\
$^{102}$ Stefan Meyer Institut f\"{u}r Subatomare Physik (SMI), Vienna, Austria\\
$^{103}$ SUBATECH, IMT Atlantique, Nantes Universit\'{e}, CNRS-IN2P3, Nantes, France\\
$^{104}$ Sungkyunkwan University, Suwon City, Republic of Korea\\
$^{105}$ Suranaree University of Technology, Nakhon Ratchasima, Thailand\\
$^{106}$ Technical University of Ko\v{s}ice, Ko\v{s}ice, Slovak Republic\\
$^{107}$ The Henryk Niewodniczanski Institute of Nuclear Physics, Polish Academy of Sciences, Cracow, Poland\\
$^{108}$ The University of Texas at Austin, Austin, Texas, United States\\
$^{109}$ Universidad Aut\'{o}noma de Sinaloa, Culiac\'{a}n, Mexico\\
$^{110}$ Universidade de S\~{a}o Paulo (USP), S\~{a}o Paulo, Brazil\\
$^{111}$ Universidade Estadual de Campinas (UNICAMP), Campinas, Brazil\\
$^{112}$ Universidade Federal do ABC, Santo Andre, Brazil\\
$^{113}$ Universitatea Nationala de Stiinta si Tehnologie Politehnica Bucuresti, Bucharest, Romania\\
$^{114}$ University of Cape Town, Cape Town, South Africa\\
$^{115}$ University of Derby, Derby, United Kingdom\\
$^{116}$ University of Houston, Houston, Texas, United States\\
$^{117}$ University of Jyv\"{a}skyl\"{a}, Jyv\"{a}skyl\"{a}, Finland\\
$^{118}$ University of Kansas, Lawrence, Kansas, United States\\
$^{119}$ University of Liverpool, Liverpool, United Kingdom\\
$^{120}$ University of Science and Technology of China, Hefei, China\\
$^{121}$ University of South-Eastern Norway, Kongsberg, Norway\\
$^{122}$ University of Tennessee, Knoxville, Tennessee, United States\\
$^{123}$ University of the Witwatersrand, Johannesburg, South Africa\\
$^{124}$ University of Tokyo, Tokyo, Japan\\
$^{125}$ University of Tsukuba, Tsukuba, Japan\\
$^{126}$ Universit\"{a}t M\"{u}nster, Institut f\"{u}r Kernphysik, M\"{u}nster, Germany\\
$^{127}$ Universit\'{e} Clermont Auvergne, CNRS/IN2P3, LPC, Clermont-Ferrand, France\\
$^{128}$ Universit\'{e} de Lyon, CNRS/IN2P3, Institut de Physique des 2 Infinis de Lyon, Lyon, France\\
$^{129}$ Universit\'{e} de Strasbourg, CNRS, IPHC UMR 7178, F-67000 Strasbourg, France, Strasbourg, France\\
$^{130}$ Universit\'{e} Paris-Saclay, Centre d'Etudes de Saclay (CEA), IRFU, D\'{e}partment de Physique Nucl\'{e}aire (DPhN), Saclay, France\\
$^{131}$ Universit\'{e}  Paris-Saclay, CNRS/IN2P3, IJCLab, Orsay, France\\
$^{132}$ Universit\`{a} degli Studi di Foggia, Foggia, Italy\\
$^{133}$ Universit\`{a} del Piemonte Orientale, Vercelli, Italy\\
$^{134}$ Universit\`{a} di Brescia, Brescia, Italy\\
$^{135}$ Variable Energy Cyclotron Centre, Homi Bhabha National Institute, Kolkata, India\\
$^{136}$ Warsaw University of Technology, Warsaw, Poland\\
$^{137}$ Wayne State University, Detroit, Michigan, United States\\
$^{138}$ Yale University, New Haven, Connecticut, United States\\
$^{139}$ Yonsei University, Seoul, Republic of Korea\\
$^{140}$  Zentrum  f\"{u}r Technologie und Transfer (ZTT), Worms, Germany\\
$^{141}$ Affiliated with an institute covered by a cooperation agreement with CERN\\
$^{142}$ Affiliated with an international laboratory covered by a cooperation agreement with CERN.\\

\end{flushleft} 

\end{document}